\documentclass[12pt,aps,superscriptaddress,prab,reprint,floatfix,nofootinbib,preprintnumbers]{revtex4-1}

\usepackage[utf8]{inputenc}
\usepackage{amssymb,amsmath}
\usepackage{graphicx}
\usepackage{subfigure}
  \usepackage{color}
\usepackage[dvipsnames]{xcolor}
 \usepackage[colorlinks]{hyperref}
  \hypersetup{
      colorlinks=true, 
      linktoc=all,     
      linkcolor=blue,  
      citecolor=magenta,
      urlcolor=blue	
  }
\usepackage{dcolumn}
\usepackage{bm}
\usepackage{interval}
\usepackage{rotating}
\usepackage{multirow}
\usepackage[separate-uncertainty = true,multi-part-units=single,multi-part-units=brackets]{siunitx}
\usepackage{url}
\usepackage{tikz}
\usetikzlibrary{arrows,decorations.markings}
\usepackage{tikz-3dplot}
\usetikzlibrary{math} 
\usetikzlibrary{calc}
\usepgflibrary{arrows}
\tikzset{
    partial ellipse/.style args={#1:#2:#3}{
        insert path={+ (#1:#3) arc (#1:#2:#3)}
    }
}

\newcommand{\dd}{\text{d}}
\def\centerarc[#1](#2)(#3:#4:#5)
    { \draw[#1] ($(#2)+({#5*cos(#3)},{#5*sin(#3)})$) arc (#3:#4:#5); }
\DeclareMathOperator{\arctantwo}{arctan2}

\definecolor{ao(english)}{rgb}{0.0, 0.5, 0.0}
\definecolor{applegreen}{rgb}{0.55, 0.71, 0.0}
\definecolor{caribbeangreen}{rgb}{0.0, 0.8, 0.6}
\definecolor{harlequin}{rgb}{0.25, 1.0, 0.0}
\begin{document}
\preprint{Draft for PR AB \today}
\title{Spin dynamics investigations for the EDM experiment at COSY}
%
%
\author{F. Rathmann}
\affiliation{Institut f\"ur Kernphysik, Forschungszentrum J\"ulich, 52425 J\"ulich, Germany}
\author{N.N. Nikolaev}
\affiliation{L.D. Landau Institute for Theoretical Physics, 142432 Chernogolovka, Russia}
\affiliation{Moscow Institute for Physics and Technology, 141700 Dolgoprudny, Russia}
\author{J.~Slim}
\affiliation{III. Physikalisches Institut B, RWTH Aachen University, 52056 Aachen, Germany}
%
%

\begin{abstract}
Precision experiments, such as the search for a deuteron electric dipole moments using a storage rings like COSY, demand for an understanding of the spin dynamics with unprecedented accuracy. In such an enterprise, numerical predictions play a crucial role for the development and later application of spin-tracking algorithms. Various measurement concepts involving polarization effects induced by an RF Wien filter and static solenoids in COSY are discussed. The matrix formalism, applied here, deals \textit{solely} with spin rotations \textit{on the closed orbit} of the machine, and is intended to provide \textit{numerical} guidance for the development of beam and spin-tracking codes for rings that employ realistic descriptions of the electric and magnetic bending and focusing elements, solenoids etc., and a realistically-modeled RF Wien filter. 
\end{abstract}
\maketitle
\tableofcontents

\section{Introduction}
The Standard Model (SM) of Particle Physics is not capable to account for the apparent matter-antimatter asymmetry of the Universe. Physics beyond the SM is required and it is either probed by employing high energies (\textit{e.g.}, at LHC), or by striving for ultimate precision and sensitivity (\textit{e.g.}, in the search for electric dipole moments). Permanent electric dipole moments (EDMs) of particles violate both time reversal $(\mathcal{T})$ and parity $(\mathcal{P})$ invariance, and are via the $\mathcal{CPT}$-theorem also $\mathcal{CP}$-violating. Finding an EDM would be a strong indication for physics beyond the SM, and pushing upper limits further provides crucial tests for any corresponding theo\-retical model,\textit{ e.g.}, SUSY. 

Up to now, EDM searches mostly focused on neutral systems (neutrons, atoms, and molecules). Storage rings, however, offer the possibility to measure EDMs of charged particles by observing the influence of the EDM on the spin motion in the ring. These \textit{direct} searches of \textit{e.g.}, proton and deuteron EDMs  bear the potential to reach sensitivities beyond $\SI{e-29}{e.cm}$. Since the Cooler Synchrotron COSY\footnote{\textcolor{blue}{The synchrotron and storage ring COSY accelerates and stores unpolarized and polarized proton or deuteron beams in the momentum range of 0.3 to \SI{3.65}{GeV/c}\,\cite{PhysRevSTAB.18.020101}.}} at the Forschungszentrum J\"ulich provides polarized protons and deuterons up to momenta of 3.7 GeV/c, it constitutes an ideal testing ground and a starting point for such an experimental program.

The investigations presented here, carried out in the framework of the JEDI Collaboration\footnote{J\"ulich Electric Dipole moment Investigations\,\cite{jedi-collaboration}.}, are relevant for the preparation of the deuteron EDM measurement\,\cite{Rathmann:2019vtb}. A radio-frequency (RF) Wien filter (WF)\,\cite{Slim:2016pim,Slim:2016dct,Slim:2017bic} makes it possible to carry out EDM measurements in a conventional \textit{magnetic} machine like COSY. The idea is to look for an EDM-driven resonant rotation of the deuteron spins from the horizontal to vertical direction and vice versa, generated by the RF Wien filter at the spin precession frequency. The RF Wien filter \textit{per se} is transparent to the EDM of the particles, its net effect is a frequency modulation of the spin tune, the number of spin precessions per turn. This modulation couples to the EDM precession in the static motional electric field of the ring, and generates an EDM-driven up-down oscillation of the beam polarization\,\cite{PhysRevSTAB.16.114001}. 

The search for EDMs of protons, deuterons, and heavier nuclei using storage rings\,\cite{srEDM-collaboration,jedi-collaboration} is part of an extensive world-wide effort to push further the frontiers of precision spin dynamics of polarized particles in storage rings. In this context, the JEDI results prompted the formation of the new CPEDM collaboration\footnote{Charged Particle Electric Dipole Moment Collaboration, \url{http://pbc.web.cern.ch/edm/edm-default.htm}}, which aims at the development of a purely electric prototype storage ring, with drastically enhanced sensitivities to the EDM of protons and deuterons, compared to what is presently feasible at COSY\,\cite{1812.08535,Rathmann:2019vtb}.

Precision experiments, such as the EDM searches, demand for an understanding of the spin dynamics with unprecedented accuracy, keeping in mind that the ultimate aim is to measure EDMs with a sensitivity up to 15 orders in magnitude better than the magnetic dipole moment (MDM) of the stored particles. 

The description of the physics of the applied approach, called \textit{RF Wien filter mapping}, is discussed further in a forthcoming separate publication. The theoretical understanding of the method and its experimental exploitation are prerequisites for the planned EDM experiments at COSY\,\cite{jedi-collaboration}, and will also have an impact on the design of future dedicated EDM storage rings\,\cite{1812.08535}.

This paper discusses various polarization effects that are induced by the RF Wien filter and static solenoids in the ring. The approach taken here strongly simplifies the machine lattice, and deals \textit{solely} with spin rotations \textit{on the closed orbit}\,\cite{lee1997spin, Mane_2005}, described by the $\mathbf{SO(3)}$ formalism. One aim of the work is to obtain a basic understanding about the interplay of spin rotations in a magnetic ring equipped with an RF Wien filter and solenoid magnets, under the simplifying assumption mentioned above. In an ideal machine with perfect alignment of the magnetic elements, the spin rotations on the closed orbit are generated primarily by the dipole magnets, therefore, for the time being, spin rotations in the quadrupole magnets are not considered. 

As we shall demonstrate below, even with an idealized ring, the parametric RF resonance-driven spin rotations reveal quite a reach pattern of spin dynamics. Our results set the background for more realistic spin tracking calculations, based on recent geodetic surveys of COSY that make available position offsets, roll, and inclination parameters for the quadrupole and dipole magnets. The treatment of the spin transport through these individually misaligned magnetic elements, can, however, be readily incorporated in the applied matrix formalism. Besides that, the spin dynamics simulations carried out in the framework of the present paper, will serve as a valuable crosscheck of the analytic approximate treatment of the parametric spin resonance, based on the Bogolyubov-Krylov-Mitropolsky averaging technique\,\cite{Bogolyubov}.

The JEDI collaboration is presently implementing a beam-based alignment scheme at COSY, which aims at providing optimized beam-transfer properties of the quadrupole and dipole magnets in the ring, with the aim to make the beam orbit as planar as possible\,\cite{TWagner}. Once this is accomplished, the spin dynamics in the ring will be largely governed by the misaligned dipoles alone. Thus effectively,  the approach described here will appropriately describe an  EDM experiment using an RF Wien filter in a beam-based aligned ring. 

The paper is organized as follows. In Sec.\,\ref{sec:spin-rotations-in-the-ring}, the effect of an EDM on the spin-evolution in a ring is discussed in terms of the Thomas-BMT equation. The inclusion of an RF Wien filter in an otherwise ideal ring is treated in Sec.\,\ref{sec:RF-Wien-filter-plus-ring}, while the polarization evolution with an RF Wien filter and additional solenoids is discussed in Sec.\,\ref{sec:polarization-evolution-with-RF-Wien-filter-and-solenoids}. The main findings are summarized in the conclusions in Sec.\,\ref{sec:conclusions}. A brief outlook into additional aspects planned to be investigated using the simulation approach taken here in the near future is also given.

\section{Spin rotations in the ring}
\label{sec:spin-rotations-in-the-ring}
\subsection{Thomas-BMT equation}

Below, the basic formalism to decribe the spin evolution in electric and magnetic fields is briefly reiterated. The generalized form of the Thomas-BMT equation describes the spin motion of a particle with spin $\vec S$ in an arbitrary electric ($\vec E$)  and magnetic field ($\vec B$). Including EDMs (in SI units), it reads\,\cite{Fukuyama:2013ioa},
\begin{equation}
 \frac{\dd \vec S}{\dd t} =  \underbrace{\left(\vec \Omega^\text{MDM} + \vec \Omega^\text{EDM} \right) }_{ \displaystyle = \vec \Omega^\text{tot}} \times \vec S\,,
 \label{eq:BMT-EDM-MDM}
\end{equation}
where
\begin{widetext}
\begin{equation}
 \begin{split}
 \vec \Omega^\text{MDM} & = -\frac{q}{m} \left[ \left( G + \frac{1}{\gamma} \right) \vec B - \frac{G \gamma}{\gamma +1} \left(\vec \beta \cdot \vec B \right)\vec \beta - \left(G + \frac{1}{\gamma +1} \right) \frac{\vec \beta \times \vec E}{c}\right]\,, \\
 \vec \Omega^\text{EDM} & = -\frac{q}{mc} \frac{\eta_\text{EDM}}{2} \left[\vec E - \frac{\gamma}{\gamma +1} \left(\vec \beta \cdot \vec E \right)\vec \beta + c\vec \beta \times \vec B \right]\,.
 \label{eq:OmegaMDM-EDM}
 \end{split}
\end{equation}
\end{widetext}
Here $m$, $\gamma$, and $\beta$ are the mass, Lorentz factor, and the velocity of a particle in units of the speed of light $c$ in vacuum, $\vec S$ is given in the particle rest frame, and the fields $\vec E$ and $\vec B$ are in the laboratory system. The magnetic dipole moment $\vec \mu$ (MDM) and the electric dipole moment $\vec d$ (EDM) are defined via the dimensionless Land\'e-factor $g$ and $\eta_\text{EDM}$
\begin{equation}
  \vec \mu = g \frac{q}{2m} \vec S, \quad \text{and} \quad \vec d = \eta_\text{EDM} \frac{q}{2m\,c} \vec S\,, 
 \label{eq:defninitions-eta-mu}
\end{equation} 
and the magnetic anomaly is given by
\begin{equation}
 G = \frac{g-2}{2}\,.
\end{equation}


\begin{table*}[htb]
\renewcommand{\arraystretch}{1.25}
\caption{\label{tab:list-of-parameters} Parameters of the deuteron kinematics, the  COSY ring, the deuteron elementary quantities, the electric dipole moment (EDM) assumed, and the field integrals of the idealized RF Wien filter (to eight decimal places). The deuteron momentum $P$ is used to specify the deuteron kinetic energy $T$, and the Lorentz factors $\beta$ and $\gamma$. The COSY circumference $\ell_\text{COSY}$ is used to specify the COSY revolution frequency $f_\text{rev}$ and the spin-precession frequency $f_s$. The deuteron mass $m$ and the deuteron  $g$ factor, taken from the NIST database\,\cite{nist} (not from the most recent one), are used to specify $G$. The deuteron EDM $d$ is used to quantify $\eta_\text{EDM}$ and $\xi_\text{EDM}$.}
\begin{ruledtabular}
 \begin{tabular}{lll}
  \textbf{Quantity}                	 	&                                    	& \textbf{Value} \\ \hline
  deuteron momentum (lab)          		& $P$                                	& \SI{970.00000000}{MeV \per c} \\
  deuteron energy (lab)            		& $T$                                	& \SI{235.97981668}{MeV} \\
  Lorentz factor (lab)				& $\beta$			     	& \num{0.45936891} \\
  Lorentz factor (lab)				& $\gamma$				& \num{1.12581478} \\ \hline
  COSY circumference         			& $\ell_\text{COSY}$                 	& \SI{183.57200000}{m} \\
  COSY revolution frequency  			& $f_\text{rev}$                     	& \SI{750197.93487176}{Hz} \\
  COSY spin precession frequency		& $f_s$					& \SI{120764.75147311}{Hz} \\ \hline
  deuteron mass	             			& $m$                                	& \SI{1875.61279300}{MeV} \\			
  deuteron $g$ factor		         	& $g$				     	& \num{1.71402546}\\
  deuteron $G = (g-2)/2$			& $G$			             	& \num{-0.14298727} \\ \hline		
  deuteron EDM                			& $d$                                	& \SI{e-20}{e.cm}\\
  deuteron dimensionless $\eta_\text{EDM}$	
						& $\eta_\text{EDM}$  			& \num{1.90102028e-06} \\
  deuteron EDM tilt angle			& $\xi_\text{EDM}$		     	& \num{-3.05366207e-06} \\	\hline	
  RF Wien filter field amplification factor  	& $f_\text{ampl}$			& \num{e3} \\	
  RF Wien filter electric field integral 	& $\int E^\text{WF}_x \dd z$		& \SI{2.20000000e+06}{V} \\ 
  RF Wien filter magnetic field integral	& $\int B^\text{WF}_y \dd z$		& \SI{1.59749820e-02}{T.m} \\ 
   RF Wien filter length 			& $\ell_\text{WF}$			& \SI{1.55000000}{m}
 \end{tabular}
\end{ruledtabular}
\end{table*}

\subsection{EDM tilt angle $\xi$ from Thomas-BMT-equation}
In an ideal machine without unwanted magnetic fields, the axis about which the particle spins precess is given by the purely vertical magnetic field $\vec  B =\vec B_\perp = B_\perp \cdot \vec e_y$. Equating the  COSY angular orbit frequency $\Omega_\text{rev} = 2 \pi f_\text{rev}$ and the relativistic cyclotron angular frequency 
\begin{equation}
\begin{split}
 \vec \Omega_\text{rev} & =  \left(
 \begin{array}{c}
   0 \\
   2\pi\cdot f_\text{rev} \\
   0
 \end{array}
 \right) = 
 \vec \Omega_\text{cyc} \\
 & = -\frac{q}{\gamma \, m} \left( B_\perp - \frac{\vec \beta \times \vec E}{\beta^2 c} \right)\,,
 \label{eq:cyclotron-frequency}
\end{split}
\end{equation}
yields, for $\vec E = 0$ with the parameters given in Table\,\ref{tab:list-of-parameters}, a vertical magnetic field of
\begin{equation}
 \vec B_\perp = 
 \left( 
 \begin{array}{c}
  0 \\
  \num{1.1075e-01}\\
  0
 \end{array}
 \right)
 \, \si{T}\,,
\label{eq:Bfromcyclotron}
 \end{equation}
which can be considered as the field that corresponds to an equivalent COSY ring where the magnetic fields are evenly distributed.

Inserting $\vec B$ from Eq.\,(\ref{eq:Bfromcyclotron}) and $\vec E = 0$ into Eq.\,(\ref{eq:OmegaMDM-EDM}), yields for the angular frequencies in the particle rest system
\begin{equation}
\begin{split}
    \vec \Omega^\text{tot} & = \vec \Omega^\text{MDM} + \vec \Omega^\text{EDM} = 
    -\frac{q}{m}
    \left(
    \begin{array}{c}
       \frac{1}{2}\eta_\text{EDM} \beta \\
       G + \frac{1}{\gamma} \\
       0
    \end{array}
    \right) B_\perp \\
    & =
 \left(
 \begin{array}{c}
       \num{-2.3171} \\
      \num{3954845.3298} \\
       \num{0.0000}      
 \end{array} 
 \right)\, \si{\per \second}.
\end{split}
\end{equation}
In the laboratory system, however, we observe with the parameters of Table\,\ref{tab:list-of-parameters} the precession frequency with respect to the cyclotron motion of the momentum,
\begin{equation}
\begin{split}
 \vec \Omega^\text{Lab} & = \vec \Omega^\text{tot} - \vec \Omega_\text{rev} = 
 -\frac{q}{m} 
 \left(
 \begin{array}{c}
  \frac{1}{2}\eta_\text{EDM} \beta \\
  G \\
 0
 \end{array}
 \right)B_\perp \\
 & = 
  \left(
 \begin{array}{c}
   \num{-2.3171} \\
   \num{-758787.3121} \\
   \num{0.0000}
 \end{array}
 \right) \, \si{\per \second}
 \,,
 \label{eq:omega-tot-lab}
\end{split} 
\end{equation}
where $\vec \Omega_\text{rev}$ denotes the COSY angular frequency along $\vec e_y$.
The spin-precession frequency yields the familiar value of
\begin{equation}
 \frac{\vec \Omega^\text{Lab}}{2 \,\pi} =   
 \left(
 \begin{array}{c}
   \num{-0.3688} \\
   \num{-120764.7515} \\
   \num{0.0000}
 \end{array}
 \right) \, \si{\per \second}\,,
\end{equation}
which is also listed in Table\,\ref{tab:list-of-parameters}. The angle by which the stable spin axis is tilted, \textit{i.e.}, the angle between $\vec \Omega^\text{Lab}$ and $\vec e_y$ is obtained by evaluating
\begin{equation}
 \xi =  \arctan \left| \frac{\vec \Omega^\text{Lab} \times \vec e_y}{\vec \Omega^\text{Lab} \cdot \vec e_y} \right| \,.
 \label{eq:xiEDM2}
\end{equation}

Inspecting Eq.\,(\ref{eq:omega-tot-lab}), the effect of an EDM in a magnetic machine can be expressed by the tilt of the stable spin axis away from the vertical orientation in the ring, given by\footnote{In Eq.\,(\ref{eq:xiEDM}), an additional factor of 2 has been inserted in the denominator, correcting Eq.\,(10) of\,\cite{PhysRevAccelBeams.20.072801}). }  
\begin{equation}
 \tan  \xi_\text{EDM} = \frac{\eta_\text{EDM} \, \beta }{2 G}\,.
\label{eq:xiEDM}
\end{equation}
For an assumed EDM of $d = \SI {1e-20}{e.cm}$, and for deuterons at a momentum of \SI{970}{MeV \per c}, Eqs.\,(\ref{eq:defninitions-eta-mu}) and (\ref{eq:xiEDM}) yield $\xi_\text{EDM}$ and $\eta_\text{EDM}$, as listed in Table\,\ref{tab:list-of-parameters}.


\subsection{Rotation matrices}
\label{sec:rotation-matrices}
Our description of the spin dynamics is based on the $\mathbf{SO(3)}$ formalism. A rotation by an angle $\theta$ around an arbitrary axis given by the unit vector $\vec n = (n_1, n_2, n_3)$ is described by the matrix\,\cite{MATHEDUC.06151274}
\begin{equation}
\mathbf{R}(\vec n, \theta) =
\left( 
\begin{array}{ccc}
b_{11} & b_{12} & b_{13} \\
b_{21} & b_{22} & b_{23} \\
b_{31} & b_{32} & b_{33}
\end{array}
\right)\,,
\label{eq:generic-rotation-matrix1}
\end{equation}
with
\begin{equation}
 \begin{split}
  b_{11} & = \cos \theta + {n_1}^2  (1-\cos \theta)  \\
  b_{12} & = n_1 n_2 (1 - \cos \theta) - n_3 \sin \theta  \\
  b_{13} & = n_1 n_3 (1 - \cos \theta) + n_2  \sin \theta  \\
  b_{21} & = n_1 n_2 (1 - \cos \theta) + n_3 \sin \theta  \\
  b_{22} & = \cos \theta + {n_2}^2 (1 - \cos \theta)  \\
  b_{23} & = n_2 n_3 (1 - \cos \theta) - n_1 \sin \theta   \\
  b_{31} & = n_1 n_3 (1 - \cos \theta) - n_2 \sin \theta  \\
  b_{32} & = n_2 n_3 (1 - \cos \theta) + n_1 \sin \theta  \\
  b_{33} & = \cos \theta + {n_3}^2 (1 - \cos \theta)     \,.
 \end{split}
 \label{eq:generic-rotation-matrix2}
\end{equation}

\subsection{One turn spin rotation matrix with EDM}
With a non-vanishing EDM, in the rotation matrix of Eq.\,(\ref{eq:generic-rotation-matrix1}), the spins do not precess anymore around the vertical axis $\vec e_y $, but rather around the direction given by
\begin{equation}
\vec c\,(\xi_\text{EDM} ) = 
\left(
\begin{array}{c} 
c_1 \\
c_2 \\
c_3  
\end{array}
\right) 
= 
\left(
\begin{array}{c} 
\sin \xi_\text{EDM}  \\\
\cos \xi_\text{EDM} \\
0 
\end{array}
\right) 
\,.
\label{eq:n1-n2-n3-withEDM}
\end{equation}

Therefore, the ring rotation matrix can be obtained by inserting into Eq.(\ref{eq:generic-rotation-matrix1}) the coefficients $c_1$, $c_2$, $c_3$ from Eq.\,(\ref{eq:n1-n2-n3-withEDM}), and by setting
\begin{equation}
 \theta := \theta(t) = \omega_s \, t = 2 \pi f_s \, t\,.
 \label{eq:thetaoft}
\end{equation}
Here, the time $t$ is defined by the number of momentum revolutions $n$ in the ring,
\begin{equation}
 t = n\cdot T_\text{rev} = \frac{n}{f_\text{rev}}\,.
 \label{eq:connection-between-n-and-t}
\end{equation}

The spin-precession frequency $f_s$, related to $\vec \Omega^\text{Lab}$ introduced in Eq.\,(\ref{eq:omega-tot-lab}), can be expressed also via 
\begin{equation}
f_s =  \frac{\Omega^\text{Lab}}{2\pi} = \frac{ G  \gamma  }{ \cos \xi_\text{EDM}} \cdot  f_\text{rev} \,,
\label{eq:spin-precession-frequency}
\end{equation}
where $f_\text{rev}$ denotes the revolution frequency. A negative $G$ factor indicates that the precession proceeds opposite to the orbit revolution.

Thus, a one-turn matrix including the EDM effect is obtained by inserting  $\theta(t)$ from Eq.\,(\ref{eq:thetaoft}) into Eq.\,(\ref{eq:generic-rotation-matrix1}) at $t = T_\text{rev} = 1/f_\text{rev}$. For comparison with numerical simulations, the ring matrix is explicitly given below (to four decimal places) for the parameters listed in Table\,\ref{tab:list-of-parameters},
\begin{small}
\begin{equation}
\begin{split}
& \mathbf{U}_\text{ring}\left(\vec c,  T_\text{rev}\right) =\\
&\left( 
\begin{array}{ccc}
\num{5.3063e-01}  & \num{-1.4333e-06} & \num{-8.4760e-01} \\
\num{-1.4333e-06} & \num{1.0000e+00}  & \num{-2.5883e-06} \\
\num{8.4760e-01}  & \num{2.5883e-06}  & \num{5.3063e-01} 
\end{array}
\right)\,.
\end{split}
\end{equation}
\end{small}

\subsection{Polarization evolution in the ring}
The evolution of the polarization vector $\vec S_1$ as function of time in the ideal bare ring is then described by
\begin{equation}
 \vec S_1(t) = \mathbf{U}_\text{ring}(\vec c, t) \times \vec S_0\,,
\label{eq:polarization-evolution-without-WF}
\end{equation}
where $\vec S_0$ denotes the initial polarization vector.

\begin{figure}[tb]
\vspace{-1.5cm}
\def\roll{50}
\def\pitch{100}
\def\tilt{\pitch-90}
\def\yaw{50}
\def\xMainRot{120}
\def\zMainRot{40}
\tdplotsetmaincoords{\xMainRot}{\zMainRot}
{\centering 
\begin{tikzpicture}[scale=1.1,tdplot_main_coords,] 
\draw[very thick,-stealth] (0,0,0) -- (-4,0,0)  node[anchor=north east]{$x$};
 \draw[very thick,-stealth] (0,0,0) -- (0,0,3)  node[anchor=north east]{$y$};
 \draw[very thick,-stealth] (0,0,0) -- (0,-4,0) node[anchor=north east]{$z$ (beam)};
\begin{scope}[canvas is yx plane at z=0]
\draw[black]  (0,0) ellipse (3cm and 3cm);
\draw[magenta,-stealth, very thick]  (0,13) [partial ellipse=290:245:13cm and 13cm];
\end{scope}
\begin{scope}[canvas is yx plane at z=0]
\end{scope}
\tdplotdefinepoints(0,0,0)(-1,0,0)({-cos(\tilt)},0,{sin(\tilt)})
\tdplotdrawpolytopearc[very thick, red, -stealth]{3}{red,above, yshift=+0.35cm, xshift = 0.2cm}{$\xi_{\rm EDM}$}
\draw[blue,very thick,-stealth] (0,0,0) -- ({3*sin(\tilt)},0,{3*cos(\tilt)})  node[anchor=north west]{$y' \parallel \vec c$};
\draw[blue, thick, dotted] ({3*sin(\tilt)},0,-0.1) -- ({3*sin(\tilt)},0,{3*cos(\tilt)});
\tdplotsetrotatedcoords{0}{\pitch}{0}
\begin{scope}[tdplot_rotated_coords,canvas is yz plane at x=0]
  \draw[blue,dashed] (0,-3) -- (0,3);
  \draw[blue,dashed] (-3,0) -- (3,0);
  \draw[thick, blue]  (0,0) ellipse (3cm and 3cm);
  \draw[red,-stealth, very thick] (0,0) -- ({3*cos(70)},{3*sin(70)}) node[anchor=north, pos=0.65]{$\vec p(t)$};
\draw[blue,very thick,-stealth] (0,0,0) -- (0,-4,0)  node[anchor=north east]{$x'$};
\end{scope}
\end{tikzpicture}}
\begin{center}
\caption{\label{fig:tilt-angle-xi} The beam particles move along the $z$ direction. In the presence of an EDM, \textit{i.e.}, $\xi_\text{EDM} > 0$, the spins precess around the $\vec c$ axis, and an oscillating vertical polarization component $p_y(t)$ is generated, as shown in Fig.\,\ref{fig:only-EDM-no-WF}. }
\end{center}
\end{figure}
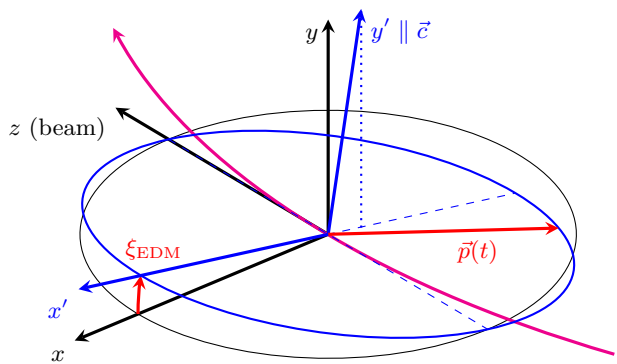

Figure\,\ref{fig:tilt-angle-xi} shows the situation when the spin rotation axis $\vec c$, defined by Eq.\,(\ref{eq:n1-n2-n3-withEDM}), is tilted with respect to the normal to the ring plane $\vec n$ ($y$-axis in the figure)\footnote{ Here, it is supposed that the polarimeter is ideally aligned to the physical ring plane so that the left-right asymmetry measures $p_y(t)$, and the up-down asymmetry measures $p_x(t)$.}. 

\begin{figure*}[htb]
\centering
\subfigure[\label{fig:only-EDM-no-WFa}  Polarization evolution of $p_x$, $p_z$ (upper panel), and $p_y$ (lower panel) for the initial spin vector $\vec S_0 = (0,0,1)$.]{\includegraphics[width=\columnwidth]{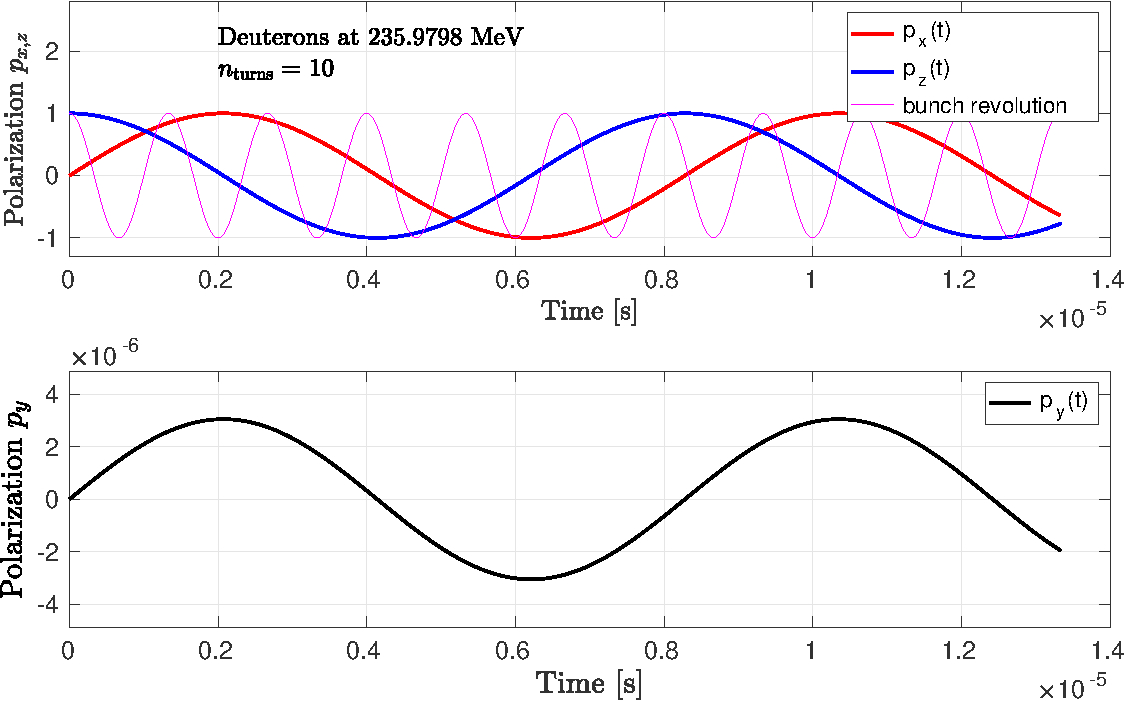}}  
\hspace{0.3cm}
\subfigure[\label{fig:only-EDM-no-WFb}  Same as panel (a), but for $\vec S_0 = (1,0,0)$.]{\includegraphics[width=\columnwidth]{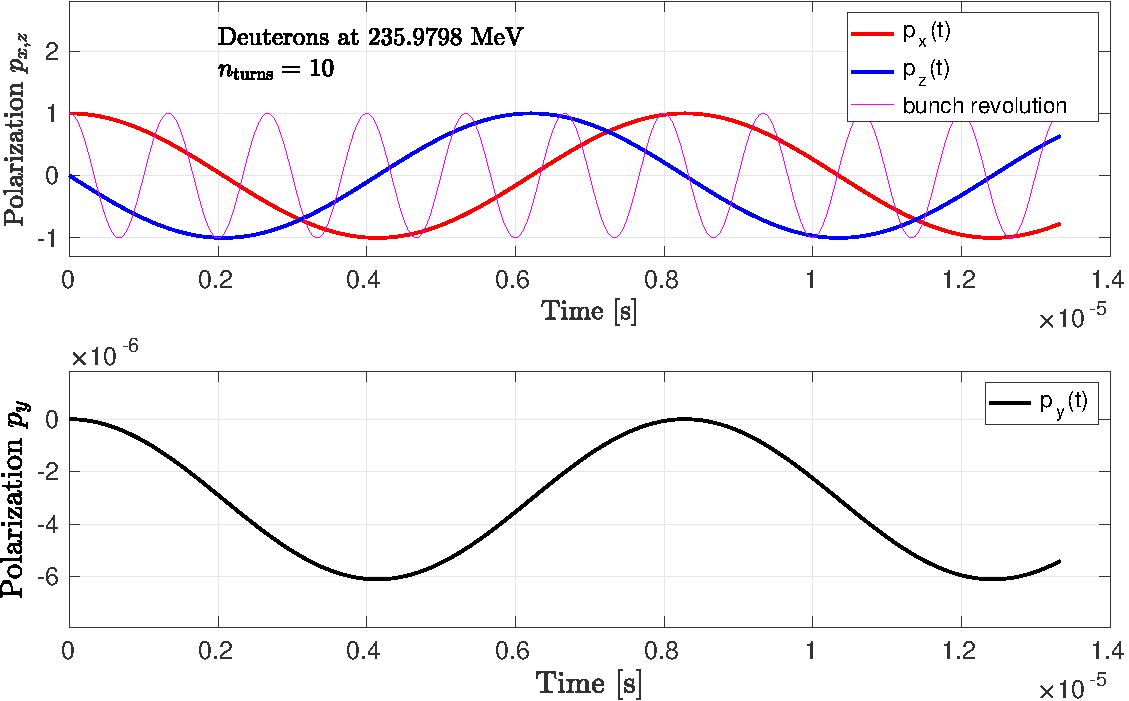}}
\caption{ \label{fig:only-EDM-no-WF} Polarization evolution during idle precession for 10 turns in an ideal ring using Eq.\,(\ref{eq:polarization-evolution-without-WF}) and the parameters listed in  Table\,\ref{tab:list-of-parameters}. Panel (a) shows $p_x(t)$, $p_z(t)$ and $p_y(t)$ for an initially longitudinal polarization, and panel.\,(b) the same for sideways polarization. The bunch revolution is indicated as well. The magnitude of the $p_y$ oscillation amplitude corresponds to the tilt angle $\xi_\text{EDM}$ (see also Eq.\,(\ref{eq:n1-n2-n3-withEDM}) and Fig.\,\ref{fig:tilt-angle-xi}).}
\end{figure*}

In Fig.\,\ref{fig:only-EDM-no-WF}, the solutions of $\vec S_1(t)$ from Eq.\,(\ref{eq:polarization-evolution-without-WF}) for two different initial in-plane polarization vectors $\vec S_0$ are shown for 10 turns. 
It is clearly visible that the polarization evolution occurs counter-clock wise with respect to the clock-wise rotation of the particles in the ring, since the deuteron $G$ factor is negative.

\section{RF Wien filter in a ring}
\label{sec:RF-Wien-filter-plus-ring}
\subsection{Electric and magnetic fields of the RF Wien filter}
The RF Wien filter, described\,\cite{Slim:2016pim}, has been designed in order to be able to manipulate the spins of the stored particles, avoiding as much as possible, the effect on the beam orbit. To this end, great care was taken to minimize the unwanted field components of the Wien filter and to characterize them via the Polynomial Chaos Expansion\,\cite{Slim201752}. In EDM mode, the main component of the magnetic induction $\vec B^\text{WF}$ is oriented along the $y$-axis, and the main component of the electric field $\vec E^\text{WF}$ along the $x$-axis. 

In order to avoid betatron oscillations in the beam, the magnetic and electric field must be matched to each other to provide a vanishing Lorentz force $\vec F_\text{L}$ (see Eq.\,(3) of \cite{Slim:2016pim}),
\begin{equation}
\vec F_\text{L} = 0  \quad \Longleftrightarrow \quad \vec E_x^\text{WF} + c \vec \beta \times \vec  B_y^\text{WF} = 0\,.
\label{eq:vanishing-Lorentz-force}
\end{equation}

According to a full-wave simulation (FWS) \footnote{CST Microwave Studio - Computer Simulation Technology AG, Darmstadt, Germany, \url{http://www.cst.com}.}, including the ferrite cage (see label 6 in Fig.\,1 of\,\cite{Slim:2016pim}), for an input power of $\SI{1}{kW}$, a field integral of $\vec B^\text{WF}$ along the beam axis of
\begin{equation}
\int_{-\ell_\text{WF}/2}^{\ell_\text{WF}/2} \vec B^\text{WF} \dd z =  
 \left(\begin{array}{ccc}
  2.73 \times 10^{-9}\\
  2.72 \times 10^{-2}\\
  6.96 \times 10^{-7}\\
 \end{array}
\right)\,\text{T\,mm}
\end{equation}
is obtained. Here, the active length of the RF Wien filter\,\cite{Slim:2016pim}, denoted by 
\begin{equation}
 \ell_\text{WF} = \SI{1550}{mm}\,,
 \label{eq:effective-length-WF}
\end{equation}
is defined as the region, where the fields are non-zero. Under these conditions, the corresponding integrated electric field components with ferrites are 
\begin{equation}
\int_{-\ell_\text{WF}/2}^{\ell_\text{WF}/2} \vec E^\text{WF} \dd z =  
 \left(\begin{array}{rrr}
  3324.577 \\
  0.018\\
  0.006\\
 \end{array}
\right)\,\text{V}\,.
\end{equation}

The design and construction of the RF Wien filter includes a ferrite cage surrounding the electrodes, which improves the field homogeneity and increases the magnitude of the fields\,\cite{Slim:2016pim}. However, in order to simplify the installation, the RF Wien filter was installed at COSY without ferrites, and in addition, it was decided to proceed without ferrites until a first direct deuteron EDM measurement is available. 

For this situation \textit{without ferrites}, and for an input power of $\SI{1}{kW}$ [ignoring the unwanted components of the field integrals ($B^\text{WF}_x$, $B^\text{WF}_z$, and $E^\text{WF}_y$, $E^\text{WF}_z$)], one obtains from the full-wave simulation (FWS) 
\begin{equation}
  \begin{split}
   \text{EDL}_x^\text{FWS} &=   \int_{-\ell_\text{WF}/2}^{\ell_\text{WF}/2} E^\text{WF}_x \dd z  \\
                           & =   \SI{2204.677323}{\volt}\,, \text{and} \\
   \text{BDL}_y^\text{FWS} & =  \int_{-\ell_\text{WF}/2}^{\ell_\text{WF}/2} B^\text{WF}_y \dd z  \\
                           & =    \SI{1.598492e-5}{\tesla \meter} \,.
  \end{split}
\end{equation}

\vspace{0.5cm}
The ratio of electric and magnetic field integrals from the FWS yields 
\begin{equation} 
\frac{1}{ \beta c} \cdot \frac{\text{EDL}_x^\text{FWS}}{\text{BDL}_y^\text{FWS}} = 1.0015 \,,
\label{eq:EB-ratio}
\end{equation}
should ideally be equal to unity.  The subsequent calculations use the field integrals of an \textit{idealized} WF with vanishing Lorentz force $\vec F_\text{L}$, given in the last column of Table~\ref{tab:WF-parameters}. 

A field amplification factor is applied in the simulations to increase the field integrals of the ideal RF Wien filter (last column Table~\ref{tab:WF-parameters}) in the simulations, so that 
\begin{equation}
\begin{split}
 \left.{\int E^\text{WF}_x \dd z}\right|_\text{used} & = f_\text{ampl} \cdot \left.{\int E^\text{WF}_x \dd z}\right|_\text{ideal} \\
 \left.{\int B^\text{WF}_y \dd z}\right|_\text{used} & = f_\text{ampl} \cdot \left.{\int B^\text{WF}_y \dd z}\right|_\text{ideal}
\end{split}
\end{equation}
The field amplification allows one to speed up the simulation calculations accordingly, \textit{without} affecting other aspects of the spin dynamics of the polarization evolution in the ring. In the description of the spin evolution via spin rotations on the closed orbit, momentum and position kicks are not considered.

\begin{table*}[htb]
\renewcommand{\arraystretch}{1.25}
\caption{\label{tab:WF-parameters}  Values for the main electric and magnetic field integrals from the full wave simulation
with and without ferrites for an input power of \SI{1}{kW} where $\vec B^\text{WF} \parallel \vec e_y$. The last column lists the electric and magnetic field integrals of an idealized Wien filter used in the simulations. In this case, the unwanted field components vanish, \textit{i.e.},  $\int E^\text{WF}_y \dd z = \int E^\text{WF}_z \dd z = \int B^\text{WF}_x \dd z =\int B^\text{WF}_z \dd z = 0$.}
\begin{ruledtabular}
\begin{tabular}{rrrr}
 Field integrals  RF Wien filter    &  $\text{with ferrites}$ & \multicolumn{2}{c}{without ferrites}  \\ 
                                    &  (real WF)     & (real WF)          &  (idealized Wien filter) \\ \hline
$\int E^\text{WF}_x \dd z$ [V]      &  $\num{3.325e3}$   & $\num{2.204677e3}$    &  $\num{2.20000000e+03}$ \\
$\int B^\text{WF}_y \dd z$ [T m]    &  $\num{2.720e-5}$  & $\num{1.598492e-5}$   &  $\num{1.59749820e-05}$ 
\end{tabular}
\end{ruledtabular}
\end{table*}

\subsection{Rotations induced by the RF Wien filter}
The effect of the RF Wien filter on the polarization evolution in the ring is implemented by an additional rotation matrix. The spin rotation in the Wien filter depends on the applied field integrals (right column of Table\,\ref{tab:WF-parameters}), multiplied by the factor $f_\text{ampl}$. 

\vspace{0.1cm}
\subsubsection{Spin rotation angle in the Wien filter}

In the following, the spin rotation angle $\psi_\text{WF}$ in the RF Wien filter is calculated numerically using the Thomas-BMT equation of Eqs.\,(\ref{eq:BMT-EDM-MDM}) and (\ref{eq:OmegaMDM-EDM}) with 
$\vec \Omega^\text{EDM} = 0$. We start with an initial spin vector
\begin{equation}
 \vec S_\text{in} = 
 \left(
 \begin{array}{c}
  0\\
  0\\
  1
 \end{array}
 \right)\,,
\end{equation}
and we compute the final polarization vector $\vec S_\text{fin}$ via
\begin{equation}
\frac{\Delta \vec S}{\Delta t} = \frac{\vec S_\text{fin} - \vec S_\text{in}}{\Delta t} 
                               =  \vec \Omega^\text{MDM} \times \vec S_\text{in}\,.
\end{equation}
Electric and magnetic field vectors for $\vec \Omega^\text{MDM}$ in Eq.\,(\ref{eq:OmegaMDM-EDM}) are obtained by computing the average fields from the idealized field integrals of the RF Wien filter (last column of Table\,\ref{tab:WF-parameters}), given by
\begin{equation}
\begin{split}
  \vec E^\text{WF}  & =
 \left(
 \begin{array}{c}
  \frac{\int E^\text{WF}_x \dd z}{ \ell_\text{WF} }\\
  0\\
  0
 \end{array}
 \right)
 \,, \text{and } \\
\vec B^\text{WF}  & =
 \left(
 \begin{array}{c}
  0\\
  \frac{ \int B^\text{WF}_y \dd z}{ \ell_\text{WF}}\\
  0
 \end{array}
 \right) 
 \,,
\end{split}
\end{equation}
where the effective length of the Wien filter is taken from Eq.\,(\ref{eq:effective-length-WF}). These conditions provide for a vanishing Lorentz force $\vec F_\text{L}$ [see also Eq.\,(\ref{eq:vanishing-Lorentz-force})].

After passing the RF Wien filter once, the final polarization vector is given by 
\begin{equation}
\begin{split}
\vec S_\text{fin} & = \left( \vec \Omega^\text{MDM} \times \vec S_\text{in} \right) \cdot \Delta t + \vec S_\text{in} \\
& \approx \left( \vec \Omega^\text{MDM} \times \vec S_\text{in} \right) \cdot \frac{\ell_\text{WF}}{\beta \, c} + \vec S_\text{in}\,,
\end{split}
\end{equation}
and, after normalizing $\vec S_\text{fin}$ to unity, the angle between $S_\text{in}$ and $\vec S_\text{fin}$ is determined from the four-quadrant inverse tangent
\begin{equation}
\arctantwo \left(  \vec S_\text{in} \times \vec S_\text{fin} , \vec S_\text{in} \cdot \vec S_\text{fin}  \right)   = 
\left(
\begin{array}{c}
 \num{0.000000} \\
 \psi_\text{WF} \\
 \num{0.000000}
\end{array}
\right) \,,
\end{equation}
with 
\begin{equation}
 |\psi_\text{WF}|  =   \SI{3.75845773e-06}{\radian}\,.
 \label{eq:psi_WF} 
\end{equation}

The spin-rotation angle in the RF Wien filter, divided by the idealized transverse magnetic field integral from Table\,\ref{tab:WF-parameters}, yields
\begin{equation}
\frac{|\psi_\text{WF}|}{\int B^\text{WF}_y \dd z} 
= \SI{2.35271485e-01}{\radian \per \tesla \per \meter}
\,.
\end{equation}


Validating the numerical result for the spin-rotation angle $\psi_\text{WF}$ in the RF Wien filter obtained in Eq.\,(\ref{eq:psi_WF}) against the analytic expression, given in \cite[Eq.\,(13)]{PhysRevAccelBeams.20.072801}, yields
\begin{equation}
\begin{split}
 \Omega_\text{WF} \cdot \Delta t =   \psi_\text{WF} 
 & =  - \frac{q}{m} \cdot \frac{(1 + G)}{\gamma^2} \cdot B^\text{WF} \cdot \frac{\ell_\text{WF}}{\beta c} \\
 & = - \frac{q}{m} \cdot \frac{(1 + G)}{\gamma^2 \beta c} \int B_\perp \dd \ell \\
 & = - \frac{q}{m} \cdot \frac{(1 + G)}{\gamma^2 \beta^2 c^2} \int E_\perp \dd \ell \\
 & = \SI{-3.75845773e-06}{\radian}  \,,
\label{eq:psi-angle-in-WF}
\end{split}
\end{equation}
where the time interval $\Delta t$ in the Wien filter has been expressed through the length $\ell_\text{WF}$.

The spin rotation angle in the RF Wien filter, given in Eq.\,(\ref{eq:psi-angle-in-WF}), constitutes an upper limit, which corresponds to a situation when a sharp $\delta$-function-like bunch passes through the device. Realistically, the bunch distribution has to be folded in, and the spin-rotation angle will be reduced correspondingly. 
\subsubsection{RF Wien filter rotation matrix}

The spin-rotation angle of the RF Wien filter changes as function of time according to
\begin{equation}
 \psi (t) = \psi_\text{WF}  \cos \left(\omega_\text{WF} \cdot t + \phi_\text{RF} \right)\,,
 \label{eq:psi-of-t-in-wien-filter-including-phase}
\end{equation}
where 
\begin{equation}
 \omega_\text{WF} = 2 \pi f_\text{WF}\,.
\label{eq:omega-of-wien-filter}
\end{equation}
The Wien filter is operated on some harmonic of the spin-precession frequency $f_s$ [Eq.\,(\ref{eq:spin-precession-frequency})], given by 
\begin{equation}
 f_\text{WF} = \left( K + \frac{G \gamma}{\cos \xi_\text{EDM}}  \right) \cdot f_\text{rev}\,, K \in \mathbb{Z} \,.
 \label{eq:WF-frequencies}
\end{equation}

The RF Wien filter rotation matrix is given by
\begin{equation}
  \mathbf{U}_\text{WF} (t) = \mathbf{R}(\vec n_\text{WF}, \psi(t))\,, 
 \label{eq:Wien-filter-matrix} 
\end{equation}
where in the generic case, $\vec n_\text{WF}$ is a unit vector along the magnetic field of the Wien filter. The case
\begin{equation}
\vec n_\text{WF} = \vec e_y\,,
\label{eq:definition-of-EDM-mode}
\end{equation}
for instance, denotes the Wien filter EDM mode. The RF Wien filter matrix $\mathbf{U}_\text{WF} (t)$ is only evaluated once per turn when the condition 
\begin{equation}
 \mod(t,T_\text{rev}) \equiv 0
 \label{eq:mod-condition-to-evaluate-wf-once-per-turn}
\end{equation}
is met stroboscopically, otherwise, the implemented function returns the $\mathbf{I}_3$ unit matrix. 

When the Wien filter is rotated around the beam axis ($z$) by some angle $\phi_\text{rot}^\text{WF}$, and 
\begin{equation}
\renewcommand{\arraystretch}{1.25}    
\begin{split}
 \vec n_\text{WF}  & = \vec n_\text{WF}\left(\phi_\text{rot}^\text{WF}\right) = \mathbf{R}(\vec e_z, \phi_\text{rot}^\text{WF}) \times \vec e_y \\
 & = 
 \left(\begin{array}{ccc}
   \cos\left(\phi_\text{rot}^\text{WF}\right) & -\sin\left(\phi_\text{rot}^\text{WF}\right) & 0\\
   \sin\left(\phi_\text{rot}^\text{WF}\right) &  \cos\left(\phi_\text{rot}^\text{WF}\right) & 0\\
   0                     &  0                     & 1
\end{array}\right)\times \vec e_y\,,
\end{split}
\label{eq:phi-rot-wien-filter-physical-rotation}
\end{equation}
the oscillations also receive a contribution from the rotation of the MDM in the horizontal magnetic field.

\subsection{Polarization evolution in the ring with RF Wien filter}
\begin{figure}[tb]
 \centering
\resizebox{\columnwidth}{!}{
\begin{tikzpicture}[scale=1,cap=round,>=latex]
  \filldraw (0,2) circle (1pt);
  \filldraw (0,-2) circle (1pt);
  \centerarc[black,thick](0,2)(0:180:3);
  \centerarc[black,thick](0,-2)(180:360:3);
  \draw[dashed,thick] (-3,2) -- (-3,-2); 
  \draw[dashed,thick]  (3,2) -- (3,-2	); 
  \draw[very thin, gray] (-4.8,2) -- (4,2);
  \draw[very thin, gray] (-4,-2) -- (4,-2);
  \foreach \i in {1,2,...,12} {
                 \filldraw[black] ({-3*cos(172.5 - (\i-1) * 15)},{3*sin(172.5 - (\i-1) * 15)+2}) circle (2pt);
                 \draw ({-3.6*cos(7.5 + (\i-1) * 15)},{3.6*sin(7.5 + (\i-1) * 15)+2}) node {D$_{\i}$};
                 \draw[very thin, gray] (0,2) -- ({-3.2*cos((\i-1) * 15)},{3.2*sin((\i-1) * 15)+2});
          }
  \foreach \i in {13,14,...,24} {
                 \filldraw[black] ({-3*cos(172.5 - (\i-1) * 15)},{3*sin(172.5 - (\i-1) * 15) - 2}) circle (2pt);
                 \draw ({-3.6*cos(7.5 + (\i-1) * 15)},{3.6*sin(7.5 + (\i-1) * 15) - 2}) node {D$_{\i}$};
                 \draw[very thin, gray] (0,-2) -- ({-3.2*cos((\i-1) * 15)},{3.2*sin((\i-1) * 15) - 2});
          }
 \centerarc[-stealth,dashed, very thick](0,2)(180:150:4.5);
 \draw[black,very thick] (-7.0,2.1)  node[anchor=west]{$t = 0, \, T_\text{rev}$,};
 \draw[black,very thick] (-6.2,1.6)  node[anchor=west]{\ldots, $n\cdot T_\text{rev}$};
 \draw[thick] (-4.8,2) -- (-4.2,2);
 \filldraw[red] (-3,1.5) circle(3pt);
 \draw (-1.75,1.5) node {Wien filter};
 \draw (-5.3,3.9) node {$t \in (0, T_\text{rev})$};
\end{tikzpicture}}
\caption{\label{fig:sketch-ring-and-WF}
Sequence of elements in the ring, corresponding to Eq.\,(\ref{eq:polarization-evolution-with-WF}). The D$_i$ ($i = 1, \ldots, 24$) indicate the 24 dipole magnets of COSY. The counting of $t$ begins with one turn in the ring, and, as indicated, the Wien filter is passed at the end of each revolution. For the discussion presented here, the dashed lines have zero length.}
\end{figure}
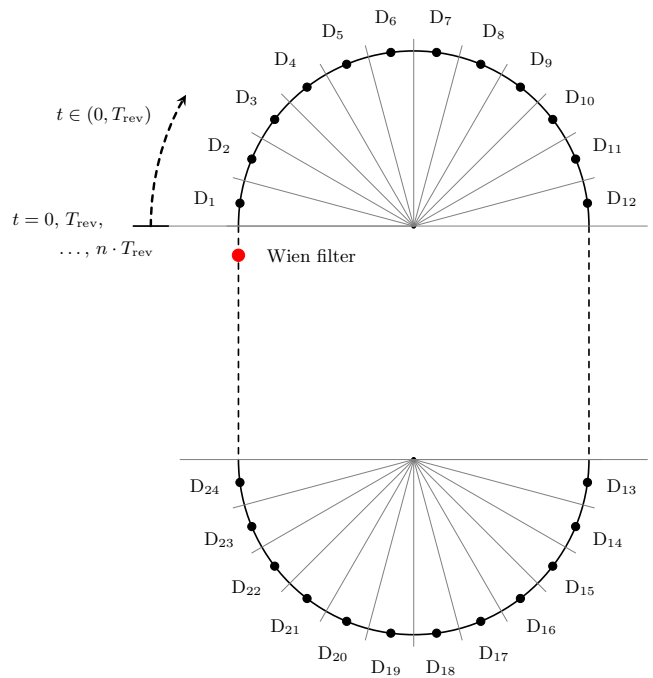

The evolution of the polarization vector $\vec S$ as function of time $t$ in the ring with RF Wien filter can be numerically evaluated via
\begin{equation}
 \begin{split}
  \vec S_2(t) = 
  & \underbrace{\mathbf{U}_\text{ring}  (\vec c, t - n\cdot T_\text{rev})}_{\text{rest of last turn}}  \\
  & \times \underbrace{\left[ \mathbf{U}_\text{WF} (t=n \cdot T_\text{rev}) \times  \mathbf{U}_\text{ring}  (\vec c, T_\text{rev}) \right]}_{\text{turn n}}  \\
  & \times\ldots  \\
  & \times\underbrace{\left[ \mathbf{U}_\text{WF} (t=2\cdot T_\text{rev}) \times  \mathbf{U}_\text{ring}  (\vec c, T_\text{rev}) \right]}_{\text{turn 2}}   \\
  & \times\underbrace{\left[ \mathbf{U}_\text{WF} (t= T_\text{rev}) \times  \mathbf{U}_\text{ring}  (\vec c, T_\text{rev}) \right]}_{\text{turn 1}} \times  \vec S_0\,.
 \end{split}
\label{eq:polarization-evolution-with-WF}
\end{equation} 
The corresponding situation is illustrated in Fig.\,\ref{fig:sketch-ring-and-WF}. The spin rotations in the ring can be described  by $\mathbf{U}_\text{ring}$. A turn begins with the revolution in the ring, and it ends with one pass through the RF Wien filter.
Between two successive points in time at which a particle encounters the RF Wien filter, its spin is just idly precessing in the machine.

According to Eq.\,(\ref{eq:polarization-evolution-with-WF}), the spin motion is stroboscopic in the sense that the spin rotation follows the angle $\psi(t)$ of the RF Wien filter [Eq.\,(\ref{eq:psi-of-t-in-wien-filter-including-phase})] turn-by-turn. The RF Wien filter therefore induces a stroboscopic turn-by-turn conversion of the transverse in-plane polarization into a vertical one (or vice versa). Using the Bogolyubov-Krylov-Mitropolsky (BKM) averaging method\,\cite{Bogolyubov}, the turn-by-turn evolution of the polarization can be approximated by the continuous dependence on the revolution number, given by $ n = f_\text{rev} \cdot t$ [Eq.\,(\ref{eq:connection-between-n-and-t})]. For the generic orientation of the RF Wien filter, the BKM averaged buildup of the vertical polarization proceeds with the resonance tune (or strength)\,\cite{PhysRevAccelBeams.20.072801}
\begin{equation}
 \varepsilon^\text{EDM} = \frac{1}{4\pi} \left|  \vec c \times \vec n_\text{WF}     \right| \cdot \psi_\text{WF}\,.
 \label{eq:resonance-tune}
\end{equation}
The direct simulations using Eq.\,(\ref{eq:polarization-evolution-with-WF}), discussed below, will furnish important crosschecks with respect to the accuracy of the analytic approximations based on the BKM averaging.

\subsection{Radial magnetic RF field in the Wien filter}
\subsubsection{Driven oscillations and resonance strength $\varepsilon^\text{\rm MDM}$}
As an illustration of the principal features of the polarization evolution, we take the case where the RF Wien filter is rotated in the so-called MDM mode with magnetic field along $-\vec e_x$, \textit{i.e.}, for $\phi_\text{rot}^\text{WF} = \SI{90}{\degree}$, where the initial polarization $\vec S_0 = -\vec e_y$.

\begin{figure}[tb]
  \centering
\includegraphics[width=1\columnwidth]{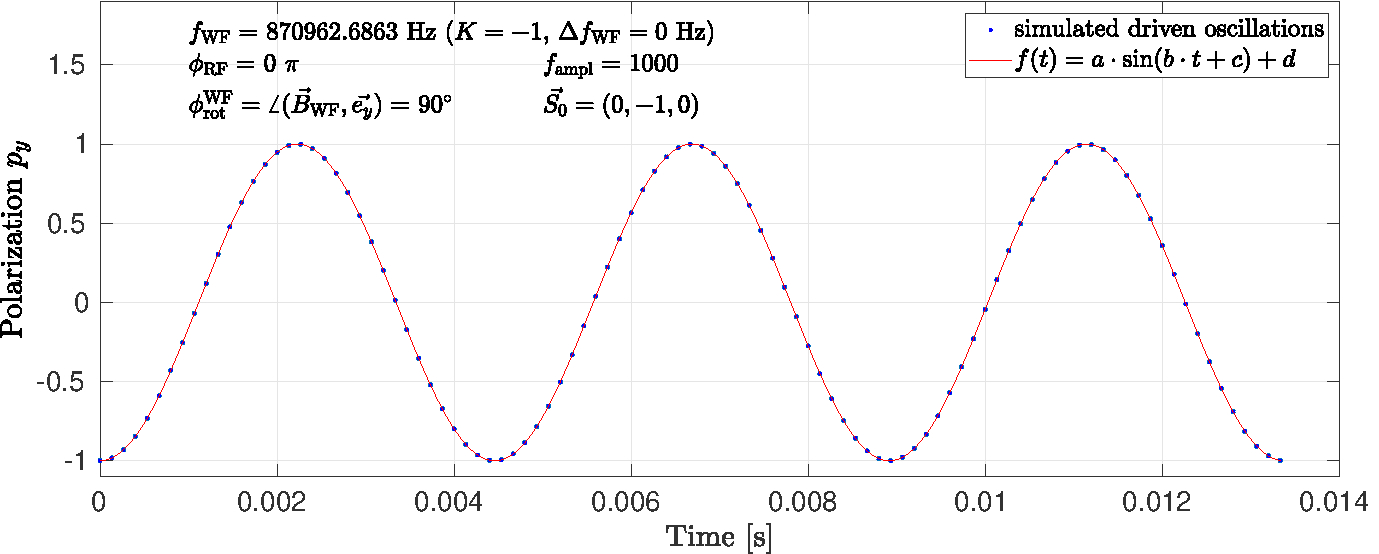}  
\caption{\label{fig:driven-oscillations} Simulated driven oscillation on resonance ($\Delta f_\text{WF} = 0$) using $\vec S_2(t)$ from Eq.\,(\ref{eq:polarization-evolution-with-WF}) with initial vertical polarization $\vec S_0 = -\vec e_y$,  $\phi_\text{RF} = 0$ [Eq.\,(\ref{eq:WF-frequencies})], and $\phi_\text{rot}^\text{WF} = \SI{90}{\degree}$ [Eq.\,(\ref{eq:phi-rot-wien-filter-physical-rotation})] for the parameters given in Table\,\ref{tab:list-of-parameters} and for the harmonic $K = -1$. The plot contains \SI{101}{points} for a total of \SI{10000}{turns}.}
 \end{figure}

Using the function for $\vec S_2(t)$, given in Eq.\,(\ref{eq:polarization-evolution-with-WF}), for the conditions of Table\,\ref{tab:list-of-parameters}, driven oscillations for RF Wien filter with magnetic field aligned along $- \vec e_x$, for $\phi_\text{rot}^\text{WF} = \SI{90}{\degree}$ [see Eq.\,(\ref{eq:phi-rot-wien-filter-physical-rotation})] were simulated. One example for $K = -1$ is shown in Fig.\,\ref{fig:driven-oscillations}. Subsequently, the simulated oscillation data were fitted using the function
\begin{equation}
 f(t) = p_y(t) = a \cdot \sin (bt + c) + d\,.
 \label{eq:fitted-function-driven-oscillations}
\end{equation}

The quality of the fit to the numerical data is evaluated in terms of squared deviations via
\begin{equation}
 \text{SSE} = \sum_{i=1}^{n_\text{points}} w_i \left[ p_y(t_i) - f(t_i)\right] ^2\,,
 \label{eq:SSE}
\end{equation}
where the weight factors are $w_i=1$, and $p_y(t) = \vec e_y \cdot \vec S_2(t)$. In the last row of Table\,\ref{tab:driven-oscillations-different-K}, the reduced $\chi^2 = \text{SSE}/\text{ndf}$ is given, where $n_\text{points} = 101$, and $\text{ndf} = n_\text{points} - 4 = 97$, since the fitted function in Eq.\,(\ref{eq:fitted-function-driven-oscillations}) has four parameters.
\begin{table}[t]
\renewcommand{\arraystretch}{1.25}   
\caption{\label{tab:driven-oscillations-different-K} Typical fit results of a simulated driven oscillation, shown in Fig.\,\ref{fig:driven-oscillations}, using \SI{10000}{turns} with  \SI{101}{data points}. The other four cases $K = \pm 1$ and $\pm 2$, within the given precision, yield identical values. $\text{SSE}/ \text{ndf}$ denotes the sum of squared deviations, computed using Eq.\,(\ref{eq:SSE}), divided by the number of degrees of freedom (ndf).}
\begin{ruledtabular}
 \begin{tabular}{rl}
 $K$                              & $-1$                                               \\ 
 $f_\text{ampl}$                  & \num{e3}                                           \\  
 $f_\text{WF}$                    & \SI{870962.6863}{Hz}                               \\ \hline 
 $a$                              & $(\num{10000        \pm 2    }) \cdot \num{e-4}$   \\
 $b$                              & \SI{1409.7817       \pm 0.0470 }{\per \second}               \\
 $c$                              & \SI{0.4997          \pm 0.0001 }{\pi}              \\
 $d$                              & (\num{0.0000        \pm 0.0001 })                  \\
 $\text{SSE}/ \text{ndf}$         & \num{3.801e-07}                                    
 \end{tabular}
\end{ruledtabular}
\end{table}

The resulting angular oscillation frequency $\Omega^\text{driven} = b$,  given in Table\,\ref{tab:driven-oscillations-different-K}, was obtained using the field integrals, listed in Table\,\ref{tab:list-of-parameters}. The uncertainties were obtained using   a computation time of about \SI{40}{s}\footnote{Lenovo T460s, all calculations use 64-bit double-precision floating point numbers, for which the $\text{machine epsilon} = \num{2.2e-16} = 2^{-52}$.}. The oscillation frequency normalized to the real magnetic field integral yields, 
\begin{equation}
 \frac{\Omega^\text{driven}}{\int B^\text{WF}_y \dd z \cdot f_\text{ampl}} = (\num{88.249} \pm 0.003) \,\, \si{\per \second \per \tesla \per \metre}\,.
\end{equation} 


The driven oscillations of the vertical polarization $p_y(t)$ (Fig.\,\ref{fig:driven-oscillations}) are induced by the horizontal magnetic field of the RF Wien filter that couples to the deuteron MDM. Since the device is operated exactly at the spin-precession frequency, the associated resonance strength or \textit{resonance tune}\,\cite{PhysRevAccelBeams.20.072801} can conveniently be expressed via
\begin{equation}
 \varepsilon^\text{MDM} = \frac{\Omega^\text{driven}}{\Omega^\text{rev} \cdot f_\text{ampl}} = \num{3e-07} \pm \num{6e-13} \,.
\end{equation}

\subsubsection{Width of the spin resonance}
\label{sec:Width-of-spin-resonance}
\begin{figure}[tb]
 \centering
 \includegraphics[width=\columnwidth]{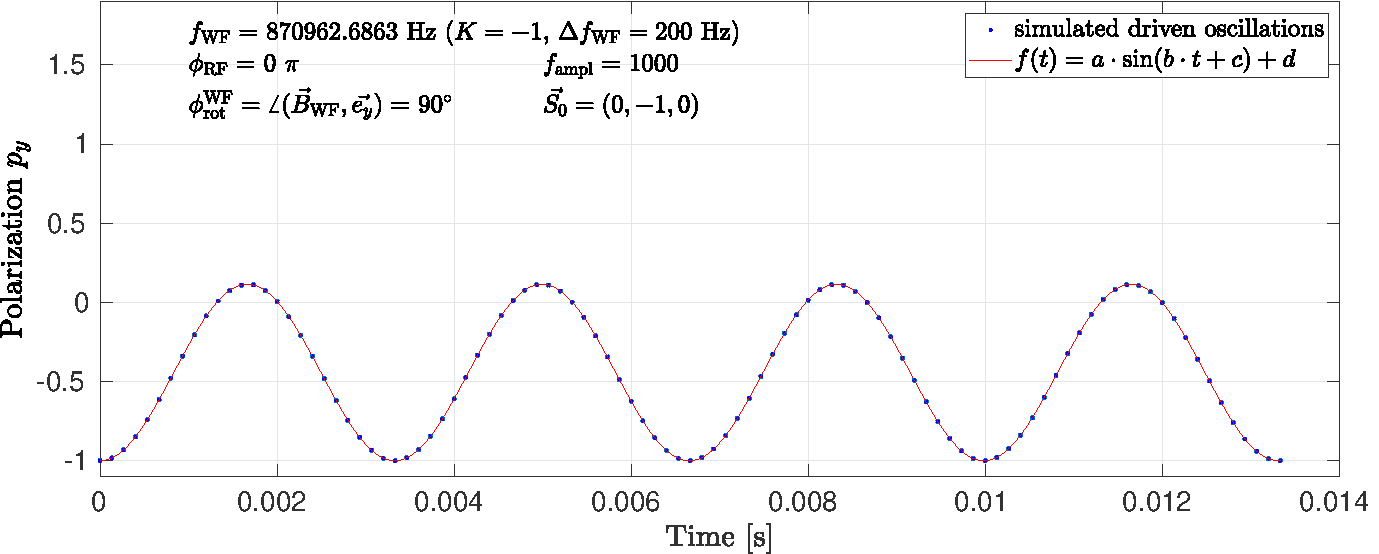}
 \caption{\label{fig:off-resonance}  Off-resonance driven oscillations with $\Delta f_\text{WF} = \SI{200}{\hertz}$ in Eq.\,(\ref{eq:off-resonance-substitution}) for the conditions of Table\,\ref{tab:list-of-parameters}.
 }
\end{figure}

The detuning of the frequency at which the RF Wien filter is operated can be parametrized by substituting in Eq.\,(\ref{eq:omega-of-wien-filter})
\begin{equation}
f_\text{WF} \rightarrow f_\text{WF} + \Delta f_\text{WF}\,.
\label{eq:off-resonance-substitution}
\end{equation}
As shown in Fig.\,\ref{fig:off-resonance}, the resulting oscillation pattern is modified. Specifically, the oscillation amplitude of $p_y(t)$ in Eq.\,(\ref{eq:fitted-function-driven-oscillations}) is altered. The argument of the sine function is subjected to the substitution
\begin{equation}
 b \cdot t =  \Omega^\text{driven} \cdot t  \rightarrow \Omega^\text{driven} \cdot\frac{ \sin(2\pi\Delta f_\text{WF} \cdot t)}{2\pi\Delta f_\text{WF}}\,,
\end{equation}
which can readily be derived from Eqs.\,(A7) and (A8) of\,\cite{PhysRevAccelBeams.20.072801}.

From a number of such simulations,  
the oscillation amplitudes and the oscillation frequencies as function of $\Delta f_\text{WF}$ are obtained by fitting. In order to reduce the time required for the simulations, again a field amplification factor of $f_\text{ampl} = \num{e3}$ was used. This leads to oscillations that are faster by the same factor. The simulated data can be described by a Lorentz curve of the form, 
\begin{equation}
L(f_\text{WF}) = \frac{a}{\left( \frac{\displaystyle \Gamma}{ \displaystyle 2} \right)^2 + (\underbrace{f_\text{WF} - f_s}_{\Delta f_\text{WF}})^2}\,.
\label{eq:Breit-Wigner-resonance}
\end{equation}

\begin{figure*}[tb]
\centering
\includegraphics[width=.75\textwidth]{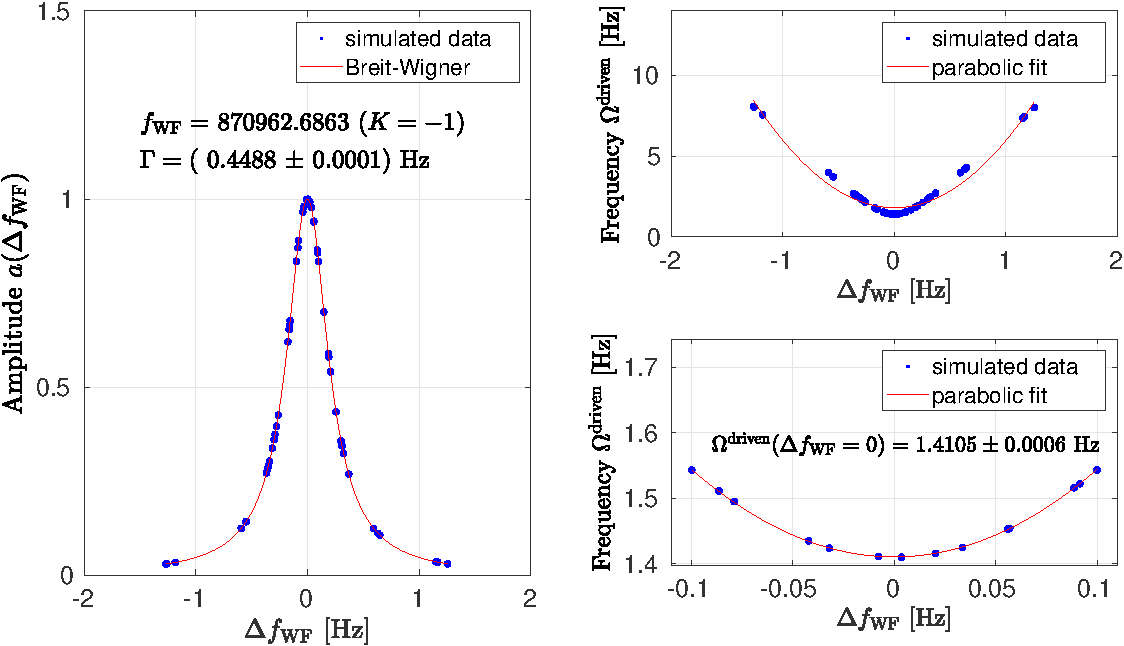}
\caption{\label{fig:resonance-width}  The left panel shows the amplitude $a$ of simulated driven oscillations as function of the frequency change $\Delta_\text{WF}$. The oscillation amplitudes were extracted from fits using Eq.\,(\ref{eq:fitted-function-driven-oscillations}). The full width at half maximum of the fitted Breit-Wigner resonance [Eq.\,(\ref{eq:Breit-Wigner-resonance})] is indicated, and the resonance curves for $K=\pm 0, 1$, and $\pm 2$ are very similar. Both panels on the right show the frequency of the driven oscillations as function of $\Delta f_\text{WF}$ together with a parabolic fit. 
}
\end{figure*}

The left panel of Fig.\,\ref{fig:resonance-width} shows the simulated spin resonance, already corrected for the field amplification factor. For all harmonic excitations used in the RF Wien filter, the simulations yield, within errors given, the same width of
\begin{equation}
   \Gamma = \SI{0.4488 \pm 0.0001}{Hz} \,.
   \label{eq:Breit-Wigner-resonance-numerical-value}
\end{equation}
Using the nominal fields of the RF Wien filter (right column of Table\,\ref{tab:WF-parameters} and $f_\text{ampl}=1$), the driven oscillations have a frequency of
\begin{equation}
   \frac{\Omega^\text{driven}}{f_\text{ampl}} = \SI{1.4105  \pm 0.0006}{Hz} \,.
\end{equation}

The two panels on the right side show that a quadratic fit to the driven oscillation frequency should be only used in a narrow region around the minimum.

The quality factor of an underdamped oscillator $Q$ is defined as
\begin{equation}
 Q = \frac{f^\text{driven}}{\Delta f^\text{driven}}\,, 
\end{equation}
where $\Delta f^\text{driven}$ is the full width at half maximum, and $f^\text{driven}$ is the resonance frequency. Thus, at a deuteron momentum of $P = \SI{970}{MeV/c}$, a theoretical estimate of the $Q$ value of the oscillating deuteron spins in the machine amounts to 
\begin{equation}
 Q = \frac{\num{120764.751}}{\num{0.4488}} \approx \num{270000}\,.
\end{equation}

\subsection{Vertical magnetic field in the RF Wien filter}
With a vertical magnetic field in the RF Wien filter ($\vec n_\text{WF} = \vec e_y$), in the expression of the spin-resonance strength [Eq.\,(\ref{eq:resonance-tune})], we then have 
\begin{equation}
 \left| \vec c \times \vec n_\text{WF}   \right| = \sin \xi_\text{EDM}\,.
\end{equation}
In this case, the experimental determination of the resonance strength $\varepsilon^\text{EDM}$ amounts to the determination of the tilt angle $\xi_\text{EDM}$ and of the associated EDM, via Eqs.\,(\ref{eq:xiEDM}) and (\ref{eq:defninitions-eta-mu}).

\subsubsection{Polarization evolution with development of $p_y(t)$}

\begin{figure}[htb]
 \centering
 \includegraphics[width = \columnwidth]{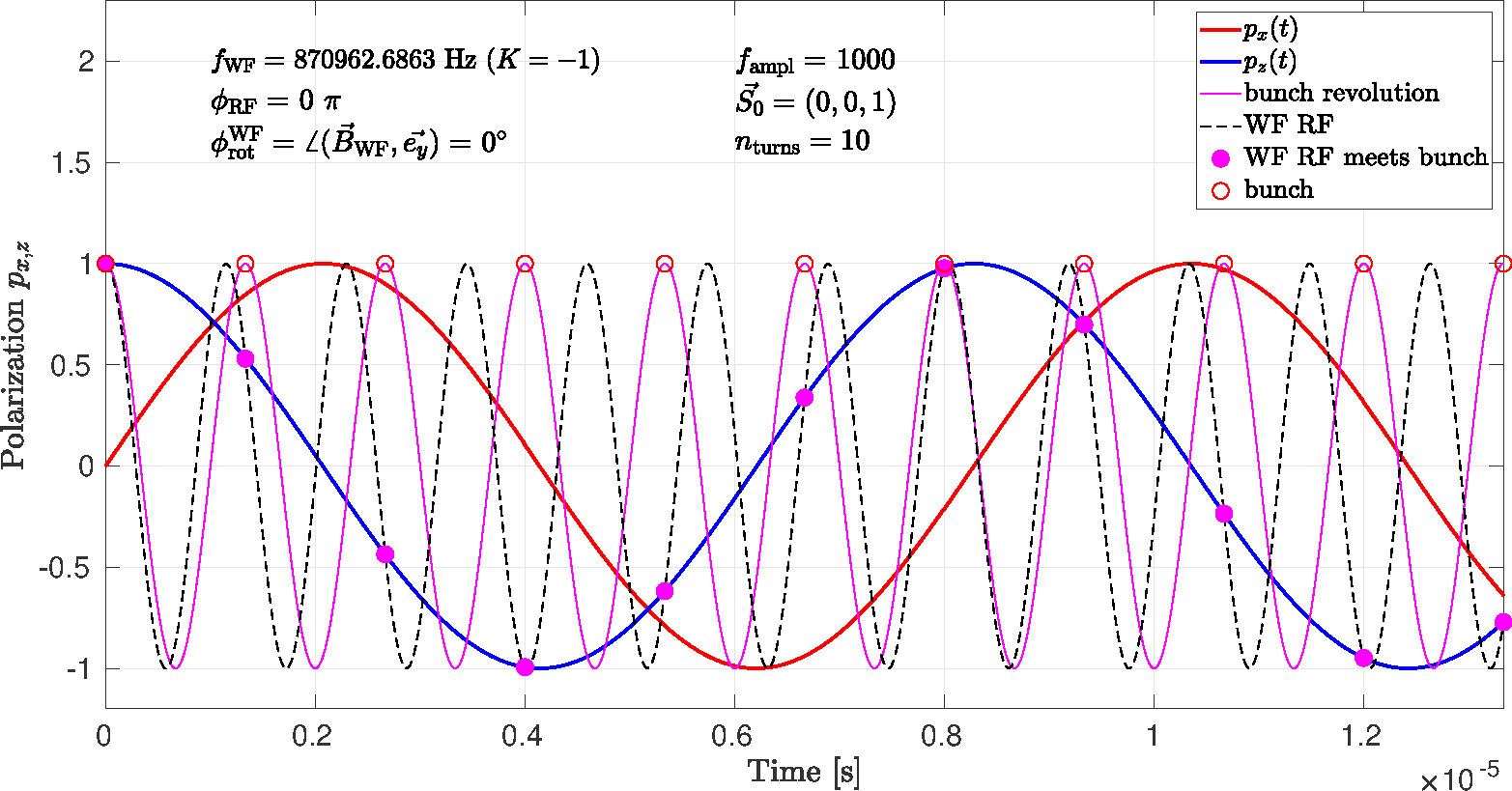}
 \caption{\label{fig:different-frequencies-with-Wien-filter} Horizontal and longitudinal polarization components $p_x(t)$ and $p_z(t)$  during the ten turns in the machine, as described by $S_2(t)$ using Eq.\,(\ref{eq:polarization-evolution-with-WF}) for the $K=-1$ harmonic and an initial polarization vector $\vec S_0$ in the horizontal ($xz$) plane. The magnetic field $\vec B^\text{WF}$ of the RF Wien filter points along $\vec e_y$, and $f_\text{ampl} = \num{e3}$. The evolution of $p_y(t)$ for the same initial condition $\vec S_0 = (0,0,1) $ is shown in Fig.\,\ref{fig:only-EDM-no-WFa}. Also indicated are the bunch revolution and the Wien filter RF frequency, and the corresponding RF amplitude when the beam bunch meets the Wien filter RF (\textcolor{magenta}{$\bullet$}).
 }
\end{figure}

In the following, the polarization buildup in the machine is addressed. The interplay of the different frequencies involved is illustrated in Fig.\,\ref{fig:different-frequencies-with-Wien-filter}.

\begin{figure}[htb]
 \centering
 \includegraphics[width = \columnwidth]{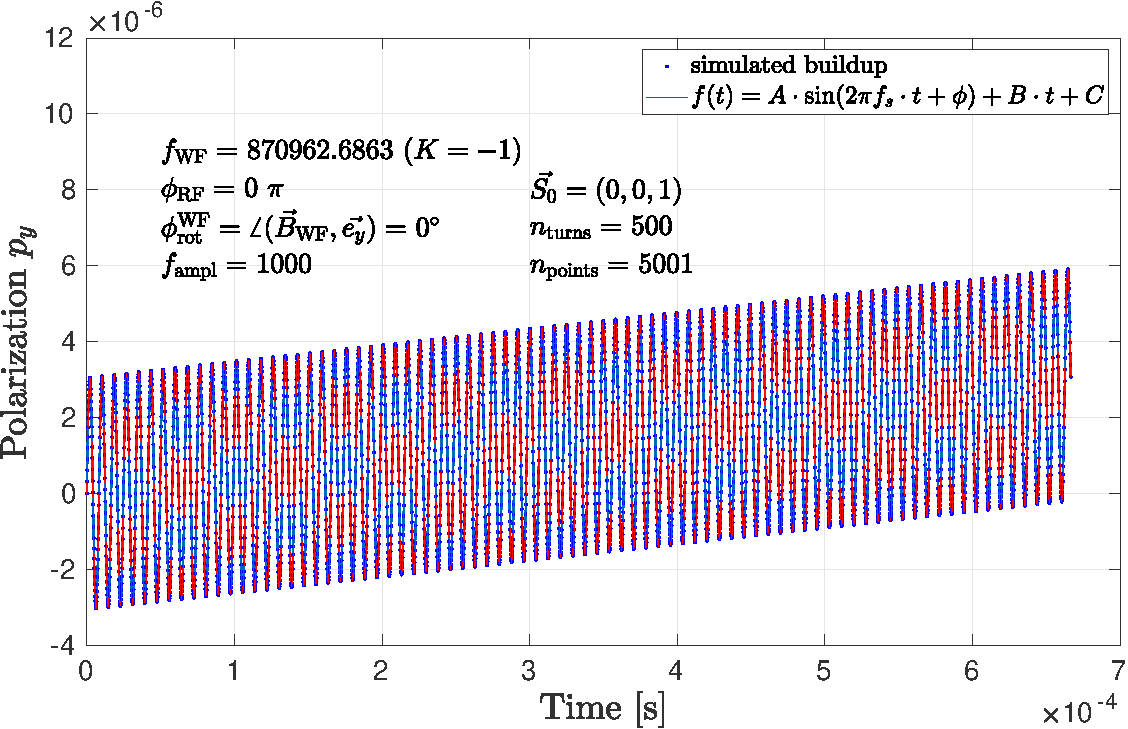}
 \caption{\label{fig:vertical-buildup-with-Wien-filter}   Buildup of a vertical polarization component for the conditions as indicated. The amplitude of the oscillating $p_y(t)$ corresponds to the EDM tilt angle $\xi_\text{EDM}$, given in Table\,\ref{tab:list-of-parameters}. The red line is a fit to the data using Eq.\,(\ref{eq:full-wave-fit}) that yields an initial slope of $\left. \dd p_y(t) / \dd t \right|_{t=0} = B = (\num{4305.059 \pm 5.268}) \times \num{e-6}\,\si{\per \second}$ (for $f_\text{ampl} = \num{e3}$).
}
\end{figure}

The same situation as in Fig.\,\ref{fig:different-frequencies-with-Wien-filter} is depicted in  Fig.\,\ref{fig:vertical-buildup-with-Wien-filter}, the only difference is the larger turn number. The graph illustrates the experimental evidence for an EDM, namely a non-vanishing slope of the vertical polarization $p_y(t)$. This slope describes the steady out-of-plane rotation of the polarization vector on the background of oscillations shown in the bottom panels of Fig.\,\ref{fig:only-EDM-no-WF}, where the oscillation amplitude $A$ perfectly matches with the angle $\xi_\text{EDM}$, used in the simulation (see Table\,\ref{tab:list-of-parameters}). 

The slope can be determined by fitting using
\begin{equation}
 p_y(t) = A \cdot \sin(2 \pi f_s \cdot t + \phi) + B \cdot t + C\,,
 \label{eq:full-wave-fit}
\end{equation}
where $f_s$ is not a fit parameter, but taken from Eq.\,(\ref{eq:spin-precession-frequency}).
Thus, using the above parametrization, the initial slope is given by
 \begin{equation}
  \left. \dot p_y(t) \right|_{t=0} = B\,.
  \label{eq:linear-slope-fit-with-offset}
 \end{equation}
 
\subsubsection{$p_y(t)$ dependence on the phases $\phi_\text{\rm RF}$ and $\phi_{S_0^x}$}
\begin{figure*}[tb]
 \centering
 \subfigure[\label{fig:definition-phi_SX}   Azimuthal angle $\phi_{S_0^x}$.]
 {
  \begin{tikzpicture}[scale=1.43,cap=round,>=latex]
    \draw (0,0) circle (4pt);
    \draw[-stealth, very thick] (1,0) -- (-3,0) node[anchor=north east]{$\vec{e}_x$};
    \draw[-stealth, very thick] (0, -1) -- (0,2) node[anchor=north east]{$\vec{e}_z$};
     \centerarc[blue,thick,-stealth](0,0)(180:140:1.5);
     \node[blue, yshift = -0.5cm] at (-2.4,1.2) {$\phi_{S_0^x} =  \angle(\vec{S_0},\vec{e}_x)$};
     \draw[red,very thick,-stealth] (0,0) -- ({2.5*cos(140)},{2.5*sin(140)})  node[anchor=west, yshift = 0.2cm, xshift = 0.2cm]{$\vec{S_0}$};
 \end{tikzpicture}
 }  
 \hspace{0.1cm}
\subfigure[\label{fig:inclination-angle-alpha}   $\vec S_y(t)$ and inclination angle $\alpha$.
]
  { \def\tilt{38.23}
 \centering
\tdplotsetmaincoords{65}{250} 
\begin{tikzpicture} [scale=6.05, tdplot_main_coords, axis/.style={->,blue,thick}, 
vector/.style={-stealth,red,very thick}, 
vector guide/.style={dashed,red,thick}]

\coordinate (O) at (0,0,0);


\pgfmathsetmacro{\ax}{0.7}
\pgfmathsetmacro{\ay}{0.6}
\pgfmathsetmacro{\az}{0.3}

\pgfmathsetmacro{\aX}{0.84}
\pgfmathsetmacro{\aY}{0.72}
\pgfmathsetmacro{\aZ}{0.0}

\coordinate (P) at (\ax,\ay,\az);

\draw[axis] (0,0,0) -- (1,0,0)     node[anchor= west, yshift = 0.2cm]{$z$};
\draw[axis] (0,0,0) -- (0,0.8,0)   node[anchor=north west]{$x$};
\draw[axis] (0,0,0) -- (0,0,0.5)   node[anchor=west]{$y$};

\draw[very thick, -stealth] (O) -- (P) node[anchor = east, yshift = 0.3cm, xshift = +1cm ]{$\vec{S}(t)$};

\draw[-stealth, very thick]   (O) -- (\ax,\ay,0) node[anchor = north, xshift = -0.4cm, yshift = 0cm ]{$\vec{S}_{xz}(t)$};

\draw[red, very thick, -stealth] (\ax,\ay,0) -- (P) node[anchor = east, yshift = -0.3cm,]{$\vec{S}_{y}(t)$};
\draw[gray, very thin] (\ax,\ay,0) -- (0,\ay,0);
\draw[gray, very thin] (\ax,\ay,0) -- (\ax,0,0);

\tdplotdefinepoints(0,0,0)(\ax, \ay, 0)(\ax, \ay, \az)
\tdplotdrawpolytopearc[very thick, red, -stealth]{0.5}{red, below, yshift=+0.5cm, xshift = -0.45cm}{$\alpha(t)$}

\tdplotdefinepoints(0,0,0)(0,0.5,0)(\ax,\ay,0)
\tdplotdrawpolytopearc[thick, -stealth]{0.4}{anchor = east, yshift=+0.1 cm, xshift = -0.4cm}{$\phi_{S^x_0}$}

\end{tikzpicture}
}
\caption{\label{fig:definition-phis}   Panel (a): Definition of the in-plane initial spin orientation angle $\phi_{S^x_0}$, and (b) relation between $\vec S_y(t)$ and the out-of-plane inclination angle $\alpha(t)$. }
\end{figure*}
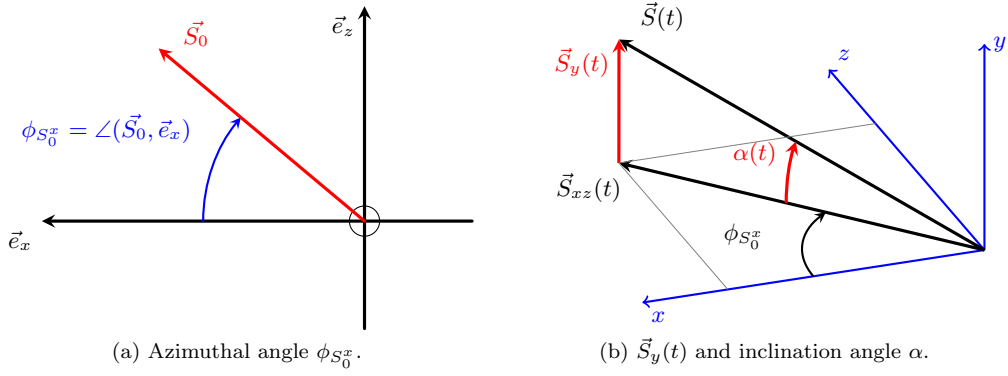

The RF phase $\phi_\text{RF}$ is introduced in Eq.\,(\ref{eq:psi-of-t-in-wien-filter-including-phase}). During a real experiment, this phase needs to be maintained by a \textit{phase-locking} system (for details see\,\cite{PhysRevLett.119.014801}). Another way to parametrize the same effect is via the angle $\phi_{S_0^x} = \angle(\vec S_0, \vec e_x)$, which is illustrated in Fig.\,\ref{fig:definition-phi_SX}.

Within the formalism described in\,\cite{PhysRevAccelBeams.20.072801}, it is the interplay between the stable spin axis $\vec c$ at the RF Wien filter and its  magnetic axis $\vec n_\text{WF}$ ($\parallel \vec B^\text{WF})$  that controls via $[\vec c \times \vec n_\text{WF}]$ the orientation of $\vec S_0$. On the other hand, one could start by fixing the orientation of $\vec S_0$ by picking some angle $\phi_{S_0^x}$. The resulting evolution of $p_y(t)$, however, must be the same, except for a possible constant shift between the two phases  $\phi_\text{RF}$ and $\phi_{S_0^x}$.

\begin{figure}[htb]
 \centering
 \includegraphics[width = \columnwidth]{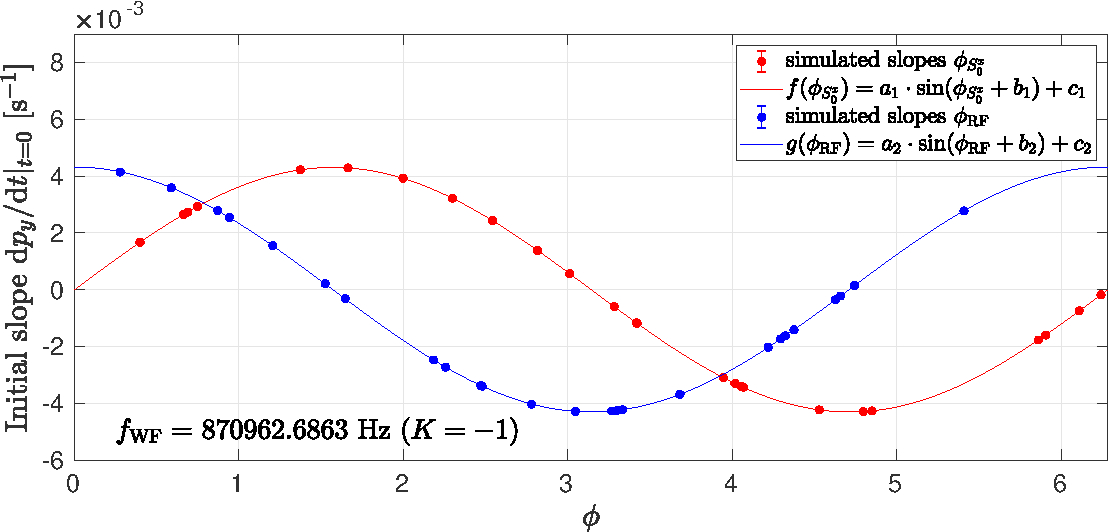}
 \caption{\label{fig:initial-slope-as-function-of-phi}
 The red (blue) curve shows the initial slope as function of 25 random values of $\phi_{S_0^x}$ ($\phi_\text{RF}$), using a field amplification factor $f_\text{ampl} = \num{e3}$. The simulated data are fitted using the functions indicated in the inset. The resulting parameters are listed in Table\,\ref{tab:summary-slopes-ideal-ring}. Each data point is obtained from a graph like the one shown in Fig.\,\ref{fig:vertical-buildup-with-Wien-filter}, but for \SI{10000}{turns} and  \SI{501}{points}. }
\end{figure}

The buildup of a vertical polarization component, which is equivalent to a rotation of the polarization vector out of the ring plane due to the EDM for a set of random azimuthal angles $\phi_{S_0^x}$ and $\phi_\text{RF}$ has been computed. The results are shown in Fig.\,\ref{fig:initial-slope-as-function-of-phi}. 
\begin{table}[tb]
\renewcommand{\arraystretch}{1.25}
 \caption{\label{tab:summary-slopes-ideal-ring}  
 Summary of parameters obtained (for $K = -1$) via fitting the oscillatory patterns of the initial slopes shown in Fig.\,\ref{fig:initial-slope-as-function-of-phi} as function of $\phi_{S_0^x}$ and $\phi_\text{RF}$, still including the factor $f_\text{ampl} =  \num{e3}$. For the other harmonics ($K=0$, $1$, and $\pm 2$), within the given uncertainties, the same values are obtained.  }
\begin{ruledtabular}
 \begin{tabular}{lll}
                           & $\phi_\text{RF}$                                   & $\phi_{S_0^x}$                         \\ \hline
 $a$                       & $(\num{4309.884   \pm 2.945}) \times \num{e-6}$    & $(\num{4304.623  \pm 2.290}) \times \num{e-6}$  \\ 
 $b$                       & $(\num{15711.584  \pm 6.254})  \times \num{e-4}$   & $(\num{-17.686   \pm 3.637}) \times \num{e-4}$ \\ 
 $c$                       & $(\num{     8.516 \pm 2.075})  \times \num{e-6}$   & $(\num{0.367     \pm 1.280}) \times \num{e-6}$  \\ 
 $\frac{\chi^2}{\text{ndf}}$       & \num{4.2e-17}                                      & \num{5.9e-17}                               
 \end{tabular}
\end{ruledtabular}
\end{table}

Within the given uncertainties, the two simulated data sets for $\phi_{S_0^x}$ and $\phi_\text{RF}$, as expected,  yield the \textit{same} results. The only difference is a phase shift of $\pi/2$ between $f(\phi_{S_0^x})$ and $g(\phi_\text{RF})$. The weights that are used to find the optimum parameters are all equal in the two data sets. 

Correcting the initial slope parameter $a$ in Table\,\ref{tab:summary-slopes-ideal-ring} for the employed field amplification factor used in the  simulation, yields a prediction for the initial slope that one would expect in an ideal ring in the presence of an EDM of $d = \SI{e-20}{e.cm}$. For an initial polarization $|\vec S_0| = 1$, with the parameters for the idealized RF WF, given in the last column of Table\,\ref{tab:WF-parameters}, one obtains
\begin{equation}
\left. \dot p_y(t) \right|_{t=0} = \frac{a(\phi_{S_0^x})}{f_\text{ampl}}  
                                                      = (\num{4.305 \pm 0.002}) \times \num{e-6} \, \si{\per \second}\,.
\label{eq:dpydt-estimate}
\end{equation}

Since the comparison of $\dot{p}_y(t) |_{t=0}$ with experiment requires knowledge of the magnitude of $\vec S(t)$, the approach taken in\,\cite{PhysRevAccelBeams.21.042002} is convenient, because the angle of the out-of-plane rotation $\alpha$ is \textit{independent} of the magnitude of the  beam polarization. The quantity of interest, indicated in Fig.\,\ref{fig:inclination-angle-alpha}, in that case is $\left. \dot\alpha (t) \right|_{t=0}$. The polarimeter measures $p_y(t)$, irrespective of the in-plane polarization $p_{xz}(t)$, given by 
\begin{equation}
 p_{xz}(t) = \sqrt{ p_{xz}(0)^2 - p_y(t)^2 }\,.
\end{equation}
From this it follows that  
\begin{equation}
\begin{split}
 \dot{\alpha}(t) & = \frac{\dd}{\dd t} \arctan\left[ \frac{p_y(t)} {p_{xz}(t)}       \right] \\
  \Rightarrow \left. \dot\alpha (t) \right|_{t=0} & = \frac{\left.\dot{p}_y(t)\right|_{t=0}} {{p}_{xz}(0)}\,.
\end{split}
\end{equation}

%

\subsubsection{Initial slope versus slow oscillation amplitude}
\begin{figure*}[tb]
\centering
\subfigure[\label{fig:initial-slopes-from-different-EDMsa}   Vertical polarization as function of time for four different EDMs for $\vec S_0 = (0,0,1)$.]{\includegraphics[width=0.47\textwidth]{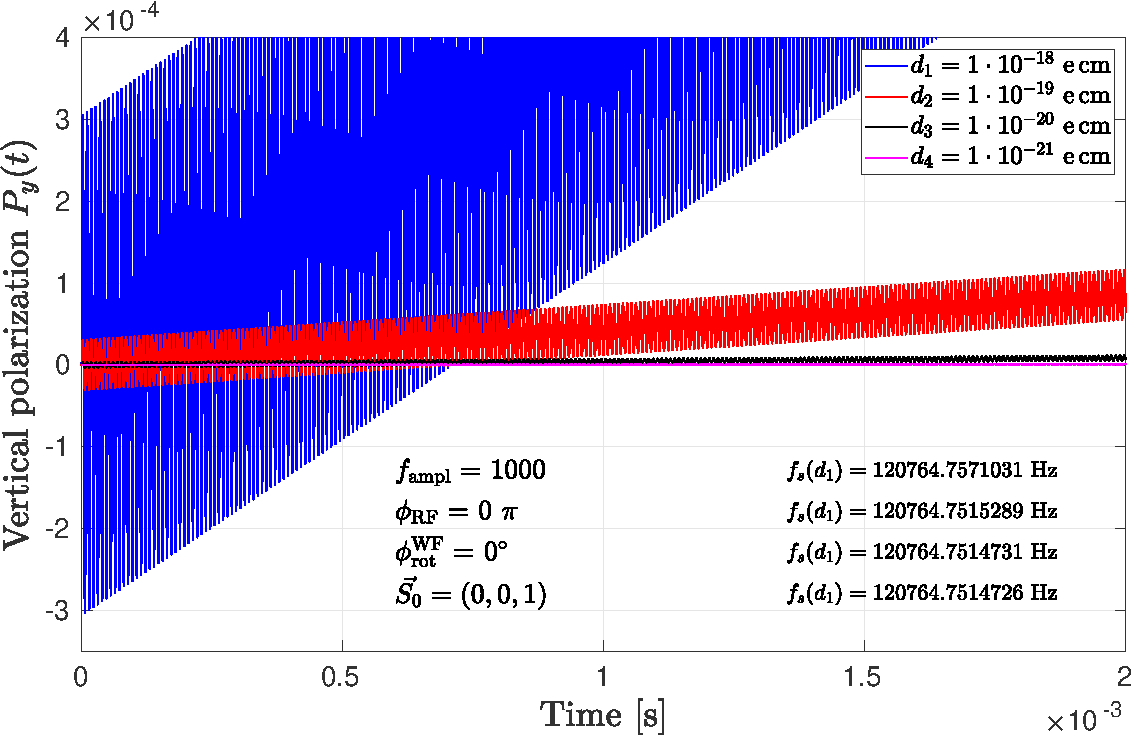}}  
\hspace{0.3cm}
\subfigure[\label{fig:initial-slopes-from-different-EDMsb}   Oscillation pattern for a large EDM and a large field amplification factor. The parametrization of the red curve in panel (b) is given in Eq.\,(\ref{eq:parametrization-of-red-curve}). ] 
{\includegraphics[width=0.49\textwidth]{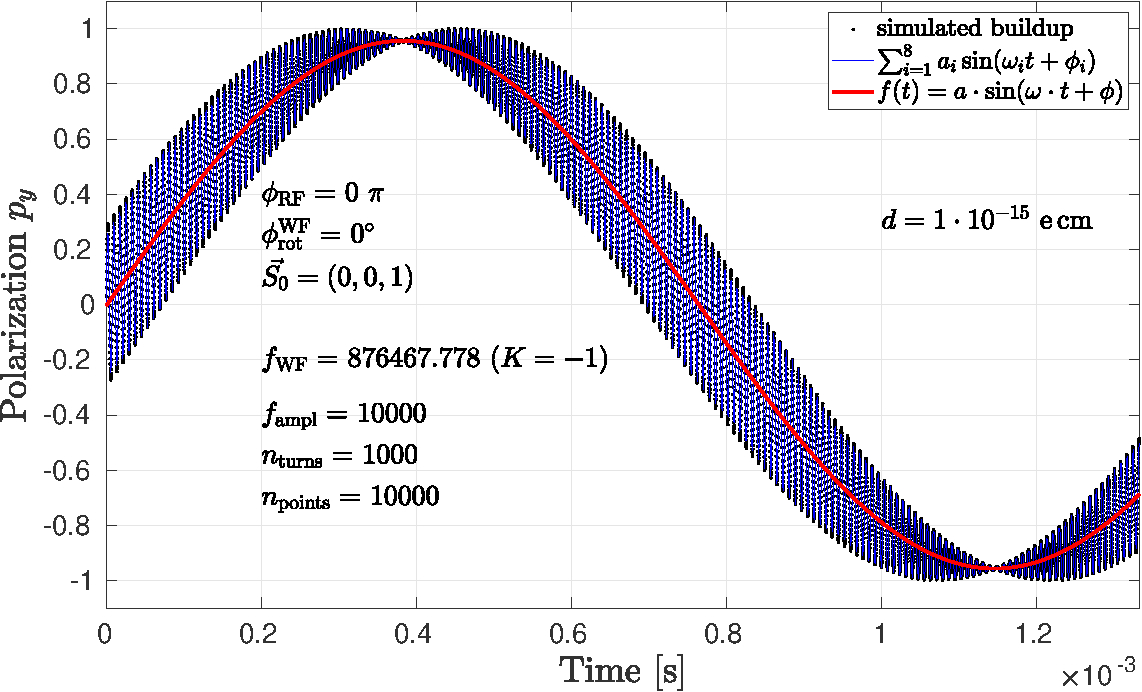}}
\caption{\label{fig:initial-slopes-from-different-EDMs} Various EDM induced oscillation pattern for short (panel a) and long evolution times (b) using different amplification factors and values for the EDM. }
\end{figure*}

Figure\,\ref{fig:initial-slopes-from-different-EDMsa} shows the initial slopes for four different assumed EDMs, for an ideal ring and an idealized Wien filter, based on the conditions listed in Table\,\ref{tab:list-of-parameters}. The EDMs manifest themselves twofold, namely in different slopes and in larger amplitudes of the fast oscillation.
The linear slopes in Fig.\,\ref{fig:initial-slopes-from-different-EDMsa} are of course just the very beginning of a sinusoidal oscillation that becomes visible only when the EDM becomes large, as depicted in Fig.\,\ref{fig:initial-slopes-from-different-EDMsb}, where
\begin{equation}
 d = \SI{e-15}{e.cm}
\end{equation}
has been used in the simulation.

The initial slope of the vertical polarization component is related to the strength of the EDM spin resonance. Another way to obtain this information is to vary the RF phase $\phi_\text{RF}$, as indicated in Fig.\,\ref{fig:initial-slope-as-function-of-phi}. The initial slope can of course also be obtained from the slow oscillation. The vertical polarization can be described by
\begin{equation}
   p_y(t) = a \sin (\omega t) \cdot  \cos \phi_\text{RF}\,,
   \label{eq:ansatz-for-py(t)}
\end{equation}
which respects the property that for any $\phi_\text{RF}$, $p_y(t)|_{t=0} = 0$. The derivative of $p_y(t)$ with respect to time is 
\begin{equation}
\begin{split}
 \dot{p}_y(t) & = a \omega \cos(\omega t) \cdot \cos \phi_\text{RF}  \\
 \Rightarrow \dot{p}_y(t)|_{t=0} & = a \omega \cdot \cos \phi_\text{RF} = (\num{3933 \pm 19})\,\si{\per \second}\,,
 \label{eq:initial-slope-from-full-oscillation}
\end{split}
\end{equation}
where the value given corresponds to the situation shown in Fig.\,\ref{fig:initial-slopes-from-different-EDMsb}. 

Numerically, the red curve in Fig.\,\ref{fig:initial-slopes-from-different-EDMsb} has been parametrized by the function 
\begin{equation}
f(t) = p_y(t) = a \sin(\omega \cdot t + \phi)\,.
\end{equation}
It turns out that the amplitude of the averaged oscillation [red curve in Fig.\,\ref{fig:initial-slopes-from-different-EDMsb}] can be determined directly from the tilt angle of the stable spin axis due to the EDM, via
\begin{equation}
 a = \cos\left( \xi_\text{EDM}(d = \SI{e-15}{e.cm}) \right) = \num{0.9564}\,.
\end{equation}
With $\xi_\text{EDM}(d = \SI{e-15}{e.cm}) = \num{-0.296373}$, within the errors, one obtains a perfect match to the value of $a$ given by 
\begin{equation}
\begin{split} 
   a      & = \num{0.9560 \pm 0.0038}\,, \\
   \omega & = \SI{4114.3813 \pm 11.8908}{\per \second}\,, \text{ and} \\
   \phi   & = \num{-0.0034 \pm 0.0082}\,.
\end{split}
\label{eq:parametrization-of-red-curve}
\end{equation}
The envelope $b^\text{osc}(t)$ of the fast oscillations is perfectly consistent with the law
\begin{equation}
b^\text{osc}(t) = \sin \xi_\text{EDM}(d) \cdot \cos (\omega t) \,.
\end{equation}


According to\,\cite{PhysRevAccelBeams.20.072801}, the EDM induced angular oscillation frequency $\omega$ in Eq.\,(\ref{eq:ansatz-for-py(t)}) can be expressed through the EDM resonance strength $\varepsilon^\text{EDM}$ and the angular revolution frequency $ \omega_\text{rev}$, via
\begin{equation}
 \omega = \varepsilon^\text{EDM} \cdot  \omega_\text{rev} 
\end{equation}
In terms of the initial slope, the resonance strength is given by
\begin{equation}
  \varepsilon^\text{EDM} = \frac{\dot{p}_y(t)|_{t=0}}{a \cos \phi_\text{RF}	} \frac{1}{ \omega_\text{rev} } \,.
  \label{eq:resaonce-strength-from-pydot}
\end{equation}
While the slopes can be easily determined as function of $\phi_\text{RF}$, the latter method using Eq.\,(\ref{eq:resaonce-strength-from-pydot}) clearly also requires knowledge about the oscillation amplitude $a$. Knowing the initial slopes alone, does not allow one to determine the resonance strength $\varepsilon^\text{EDM}$. 

\begin{figure}[tb]
 \centering
 \includegraphics[width = \columnwidth]{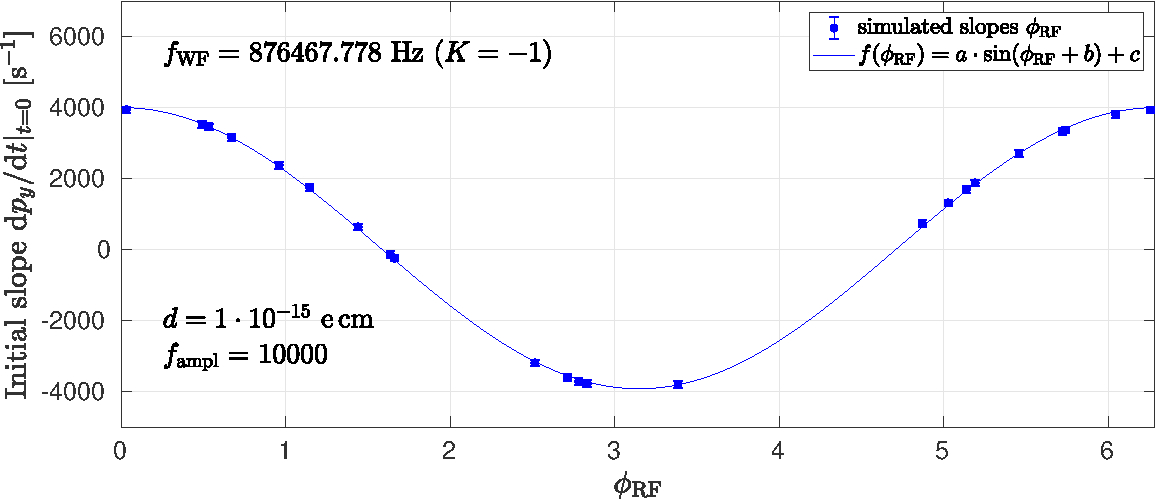}
 \caption{\label{fig:initial-slope-as-function-of-phi-large-EDM} 
 Initial slope as function of 24 random values of $\phi_\text{RF}$ using a field amplification factor $f_\text{ampl} = \num{e4}$ and the indicated EDM. The simulated data are fitted using the function indicated in the inset. The resulting parameters are $a = (\num{3959.122 \pm 35.344})$, $b = (\num{4135.901 \pm  0.009})$, and $c = (\num{39.861 \pm 25.995})$. Each data point is obtained from a graph like the one shown in Fig.\,\ref{fig:vertical-buildup-with-Wien-filter}, but for 10 turns and 1001 points.
 }
\end{figure}

Using the technique of variation of $\phi_\text{RF}$, as shown in Fig.\,\ref{fig:initial-slope-as-function-of-phi}, Fig.\,\ref{fig:initial-slope-as-function-of-phi-large-EDM} yields an initial slope of 
\begin{equation}
\left.\dot p_y(t)\right|_{t=0}  = (\num{3959 \pm  35})\,\si{\per \second}\,,
\end{equation}
which agrees numerically well within errors with the value given in the last line of Eq.\,(\ref{eq:initial-slope-from-full-oscillation}).

\subsubsection{Determination of the running spin tune, based on the polarization evolution $\vec S_2(t)$}
\label{sec:determination-of-spintune}
The standard definition of the spin tune as a rotation around the local stable spin axis $\vec n_s$ at every point in the machine does not involve a time dependence of the polarization evolution, like the one generated by the RF Wien filter. When a time-dependent polarization is involved, in the following, the term \textit{running} or \textit{instantaneous spin tune} is used. In case there is a time-dependent or instantaneous spin tune, the direction of $\vec n_s$ also changes as function of time, \textit{i.e.}, $\vec n_s \equiv \vec n_s(t)$ (see further Sec.\,\ref{sec:determination-spin-closed-orbit}).

Using the numerical simulations for $\vec S_2(t)$, or any other spin-evolution function, one can numerically determine the running spin tune in the following way. For this one needs three spin vectors from the spin-evolution function, say 
\begin{equation}
\begin{split}
 \vec a  & = \vec S_2(t)\,, \\
 \vec b  & = \vec S_2(t+T_\text{rev})\,, \text{ and } \\
 \vec c  & = S_2(t + 2\cdot T_\text{rev})
\end{split}
\end{equation}
 Using these three vectors, two more vectors  are constructed,
\begin{equation}
   \vec d(t) =  \vec a  - \vec b  \quad \text{ and } \quad   \vec e(t) =  \vec a - \vec c \,.
\end{equation}
The in-plane angle between $\vec d(t)$ and $\vec e(t)$ can be used to determine the running, time-dependent spin tune $\nu_s(t)$. To this end, we define the normal vector $\vec N$ of the plane that contains $\vec d$ and $\vec e$,
\begin{equation}
 \vec N = \frac{\vec d \times \vec e}{\left |\vec d \times \vec e \,\right|}\,,
 \label{eq:definition-of-normal-vector-N}
\end{equation}
that corresponds to the \textit{instantaneous} (running) spin axis. Using $\vec N$, we find the in-plane components of $\vec b$ and $\vec c$, via
\begin{equation}
 \vec b_\perp = \vec b \times \vec N \quad \text{ and } \quad \vec c_\perp = \vec c \times \vec N\,.
\end{equation}
The normalized versions of these vectors are called 
\begin{equation}
 \vec f = \frac{\vec b_\perp}{| \vec b_\perp |} \quad \text{ and } \quad \vec g = \frac{\vec c_\perp}{| \vec c_\perp |}\,,
\end{equation}
and the running spin tune is determined from
\begin{equation}
    \nu_s (t) = \frac{1}{2\pi} \frac{G}{|G|} \arctan \left| \frac{\vec f(t)  \times \vec g(t) }
    {\vec f(t) \cdot \vec g(t)} \right|\,.
\label{eq:spin-tune-calculated-from-S8}
\end{equation}
The factors in front of $\arctan$ take care that $\nu_s(t)$ generates the correct sign based on the $G$-factor and the number of spin-precessions per turn.
\begin{figure*}[htb]
\centering
\includegraphics[width = 1\textwidth]{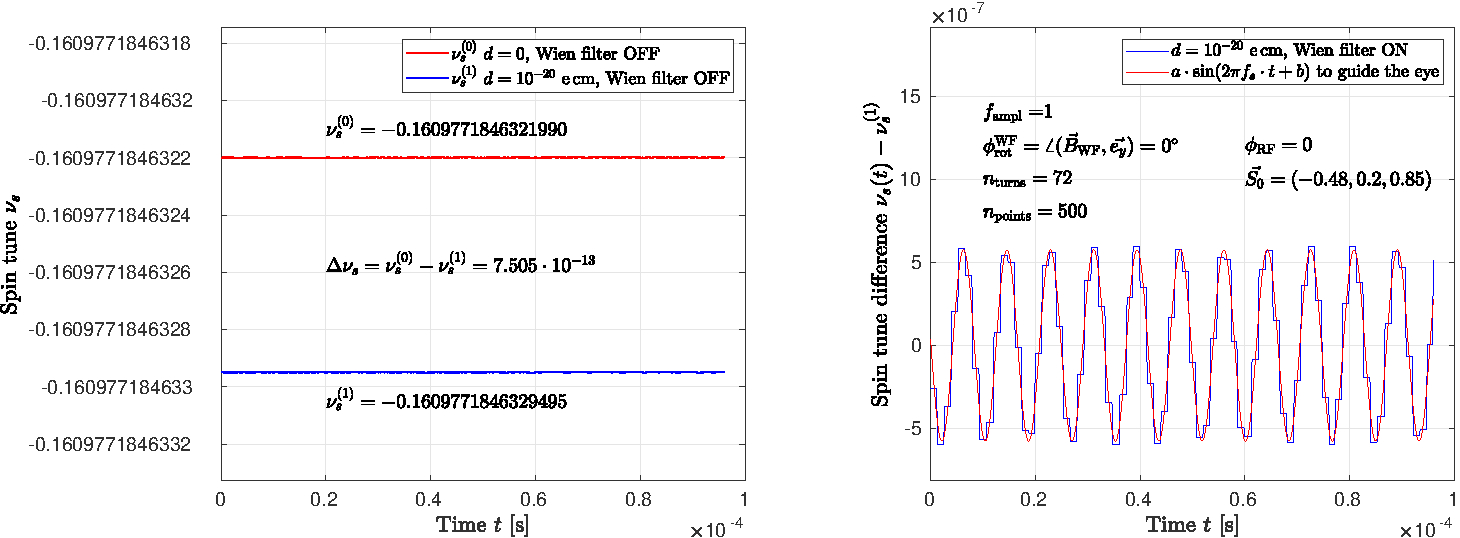}
\caption{\label{fig:spin-tunes} 
\small The graph on the \textbf{left} shows the spin tune in the machine, calculated using Eq.\,(\ref{eq:spin-tune-calculated-from-S8}) at the conditions of Table\,\ref{tab:list-of-parameters} for $d = 0$ (red) and $d = \SI{1e-20}{e.cm}$ (blue), when the RF Wien filter is switched OFF. On the \textbf{right}, the RF Wien filter is switched ON in EDM mode with $f_\text{ampl} = 1$. The red curve shows the spin oscillation frequency $f_s$ from Eq.\,(\ref{eq:spin-precession-frequency}), and the blue line the running spin tune difference $\nu_s(t) - \nu_s^{(1)}$ for each turn. It should be noted that the initial spin vector $\vec S_0$ is not in the ring ($xz$) plane (see Fig.\,\ref{fig:tilt-angle-xi}).}
\end{figure*}

As a cross check of the algorithm, with RF WF switched off, for the beam conditions given in Table\,\ref{tab:list-of-parameters}, Eq.\,(\ref{eq:spin-tune-calculated-from-S8}) yields 
\begin{widetext}
\begin{equation}
  \begin{split}
  \text{for } d = 0:                \quad \nu_s^{(0)}      = G \gamma        = & -\num{1.609771846321990e-01} \,, \\
  \text{for } d = \SI{1e-20}{e.cm}: \quad  \nu_s^{(1)}  = \frac{G \gamma}{\cos\xi_\text{EDM}}   = & -\num{1.609771846329495e-01} \,,\text{and } \\
  \Delta \nu_s = \nu_s^{(0)} - \nu_s^{(1)}             = & +\num{7.505e-13}\,,
  \end{split}
\label{eq:spin-tunes-numerical-values}
\end{equation}
\end{widetext}
where all three numbers have been calculated using Eq.\,({\ref{eq:spin-tune-calculated-from-S8}). As an additional cross check, the difference of the spin tunes 
\begin{equation}
  \frac{\nu_s^{(0)}}{\cos \xi_\text{EDM}} - \nu_s^{(1)} \approx \num{e-16}\,,
\end{equation}
which is very close to the achievable machine precision\footnotemark[4].


During a revolution in the machine, as prescribed by $\vec S_2(t)$ using Eq.\,(\ref{eq:polarization-evolution-with-WF}), the spin tune remains constant during each turn (see Fig.\,\ref{fig:spin-tunes}). When the RF Wien filter is switched on, due to the additional spin rotation in the time-varying RF field, the spin tune jumps from turn to turn. The oscillation amplitude of the spin tune variation due to the RF Wien filter using a power of \SI{1}{kW} (see Table\,\ref{tab:list-of-parameters}) is well consistent with the expectation from the spin rotation formalism
\begin{equation}
 a = (\num{5.7 \pm 0.2}) \times \num{e-7} \approx \frac{|\psi_\text{WF}|}{2\pi} = \num{6.0e-7}\,.
\end{equation}
The \textit{average} spin tune, however, remains constant.

\vspace{0.1cm}
\subsubsection{Instantaneous spin orbit determination based on $\vec S_2(t)$ \label{sec:determination-spin-closed-orbit}}
The running spin orbit vector $\vec n_s$ can be easily determined from the procedure of the previous section, using the normal vector $\vec N$, defined in Eq.\,(\ref{eq:definition-of-normal-vector-N}),
\begin{equation}
 \vec n_s (t) = \vec N(t) \,.
 \label{eq:spin-closed-orbit-vector-as-fct-of-time}
 \end{equation}
Similarly to the running (instantaneous) spin tune, the instantaneous spin orbit (running spin axis) exhibits oscillatory in-plane components.
\vspace{0.1cm}

\section{Polarization evolution with RF Wien filter and solenoids}
\label{sec:polarization-evolution-with-RF-Wien-filter-and-solenoids}
\subsection{Evolution equation with additional static solenoids}

In the course of this paper, with the RF Wien filter in EDM mode ($\vec B^\text{WF} \parallel \vec e_y$),  the EDM interaction with the motional electric field in the ring, was the only source of up-down spin-oscillations.

In the following, two static solenoids in the straight sections will be added to the ring. Besides that, we shall make an allowance for rotations of the RF Wien filter around the longitudinal $\vec e_z$ (momentum) direction. Such rotations induce a radial magnetic RF field, and, in conjunction with the solenoidal magnetic fields, we start mixing the EDM and MDM induced rotations. The idea, common to all EDM experiments, is to disentangle the EDM signal from an extrapolation to a vanishing MDM contribution\,\cite{Afach:2015sja,Afach:2015ima}.

With two static solenoids added to the ring, the resulting sequence of elements is depicted in Fig.\,\ref{fig:ring-with-two-more-solenoids}.
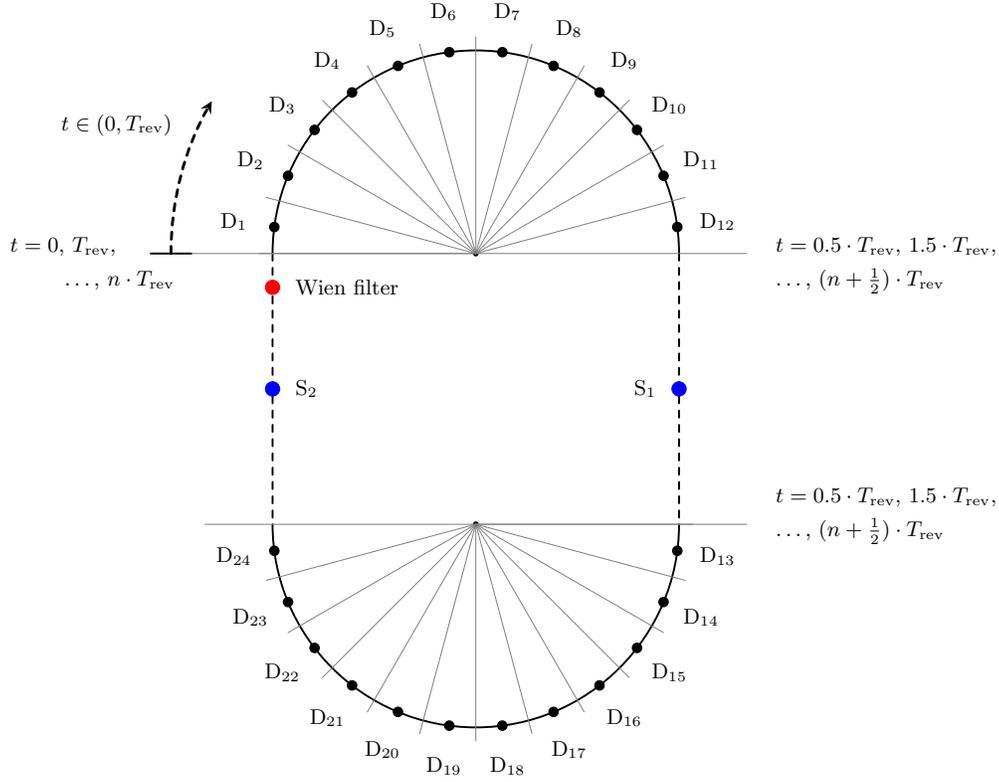
\begin{figure*}[htb]
 \centering
\resizebox{0.75\textwidth}{!}{
\begin{tikzpicture}[scale=1,cap=round,>=latex]
  \filldraw (0,2) circle (1pt);
  \filldraw (0,-2) circle (1pt);
  \centerarc[black,thick](0,2)(0:180:3);
  \centerarc[black,thick](0,-2)(180:360:3);
  \draw[dashed,thick] (-3,2) -- (-3,-2); 
  \draw[dashed,thick]  (3,2) -- (3,-2	); 
  \draw[very thin, gray] (-4.8,2) -- (4,2);
  \draw[very thin, gray] (-4,-2) -- (4,-2);
  \foreach \i in {1,2,...,12} {
                 \filldraw[black] ({-3*cos(172.5 - (\i-1) * 15)},{3*sin(172.5 - (\i-1) * 15)+2}) circle (2pt);
                 \draw ({-3.6*cos(7.5 + (\i-1) * 15)},{3.6*sin(7.5 + (\i-1) * 15)+2}) node {D$_{\i}$};
                 \draw[very thin, gray] (0,2) -- ({-3.2*cos((\i-1) * 15)},{3.2*sin((\i-1) * 15)+2});
          }
  \foreach \i in {13,14,...,24} {
                 \filldraw[black] ({-3*cos(172.5 - (\i-1) * 15)},{3*sin(172.5 - (\i-1) * 15) - 2}) circle (2pt);
                 \draw ({-3.6*cos(7.5 + (\i-1) * 15)},{3.6*sin(7.5 + (\i-1) * 15) - 2}) node {D$_{\i}$};
                 \draw[very thin, gray] (0,-2) -- ({-3.2*cos((\i-1) * 15)},{3.2*sin((\i-1) * 15) - 2});
          }
 \centerarc[-stealth,dashed, very thick](0,2)(180:150:4.5);
 
 \draw[black,very thick] (-7.0,2.1)  node[anchor=west]{$t = 0, \, T_\text{rev}$,};
 \draw[black,very thick] (-6.2,1.6)  node[anchor=west]{\ldots, $n \cdot T_\text{rev}$};
 

 \draw[black,very thick] (4.3,2.1)  node[anchor=west]{$t = 0.5 \cdot T_\text{rev}$, $1.5 \cdot  T_\text{rev}$,};
 \draw[black,very thick] (4.3,1.6)  node[anchor=west]{\ldots, $ (n + \frac{1}{2}) \cdot T_\text{rev}$};

 \draw[black,very thick] (4.3,-1.6)  node[anchor=west]{$t = 0.5 \cdot T_\text{rev}$, $1.5 \cdot  T_\text{rev}$,};
 \draw[black,very thick] (4.3,-2.1)  node[anchor=west]{\ldots, $ (n + \frac{1}{2}) \cdot T_\text{rev}$};

 \draw[thick] (-4.8,2) -- (-4.2,2);
 \filldraw[red] (-3,1.5) circle(3pt);
 \draw (-1.9,1.5) node {Wien filter};
 \draw (-5.3,3.9) node {$t \in (0, T_\text{rev})$};
 \filldraw[blue] (-3,0) circle(3pt);
 \draw (-2.5,0) node {S$_2$};
 \filldraw[blue] (3,0) circle(3pt);
 \draw (2.5,0) node {S$_1$};
 
\end{tikzpicture}}
\caption{\label{fig:ring-with-two-more-solenoids}
Sequence of elements in the ring, corresponding to Eq.\,(\ref{eq:polarization-evolution-with-WF-and-two-solenoids}), including besides the RF Wien filter, also two static solenoids
S$_1$ and S$_2$. 
}
\end{figure*}
The one-turn ring matrix can be split into two arcs, one arc made of the dipole magnets D$_1$ to D$_{12}$, and the second arc made of dipoles D$_{13}$ to D$_{24}$. Since
\begin{equation}
 \mathbf{U}_\text{ring}\left(\vec c,  T_\text{rev}\right) = \mathbf{U}_\text{ring}^\text{arc\,2}\left(\vec c,  T_\text{rev}/2 \right) \times \mathbf{U}_\text{ring}^\text{arc\,1}\left(\vec c,  T_\text{rev}/2 \right)\,,
 \label{eq:evolution-equation-with-additional-solenoids}
\end{equation}
the two additional solenoids can be inserted before and behind arc\,2, leading to
\begin{widetext}
\begin{equation}
 \mathbf{U}_\text{ring}^\text{2\,sol}\left(\vec c,  T_\text{rev}, \chi_\text{rot}^{\text{S}_1}, \chi_\text{rot}^{\text{S}_2} \right) 
 =  \mathbf{R} \left(\vec e_z, \chi_\text{rot}^{\text{S}_2}\right) 
 \times \mathbf{U}_\text{ring}^\text{arc\,2}\left(\vec c,  T_\text{rev}/2 \right) 
 \times \mathbf{R}\left(\vec e_z, \chi_\text{rot}^{\text{S}_1}\right) 
 \times \mathbf{U}_\text{ring}^\text{arc\,1}\left(\vec c,  T_\text{rev}/2 \right)\,,
 \label{eq:}
\end{equation}
\end{widetext}
invoking again the generic rotation matrix  $\mathbf{R}(\vec e_z, \chi_\text{rot})$ from Eq.\,(\ref{eq:generic-rotation-matrix1}).

In a similar fashion as in Eq.\,(\ref{eq:polarization-evolution-with-WF}), one can write for the polarization evolution,
\begin{widetext}
\begin{equation}
 \begin{split}
  \vec S_3(t) = 
  & \underbrace{\mathbf{U}_\text{ring}  (\vec c, t - n\cdot T_\text{rev})}_{\text{rest of last turn}} \times 
   \underbrace{\left[ \mathbf{U}_\text{WF} (t=n \cdot T_\text{rev}) \times  \mathbf{U}_\text{ring}^\text{2\,sol}  \left(\vec c, T_\text{rev},\chi_\text{rot}^{\text{S}_1}, \chi_\text{rot}^{\text{S}_2} \right) \right]}_{\text{turn n}}  \\
  & \times  \ldots \\
  & \times \underbrace{\left[ \mathbf{U}_\text{WF} (t=2\cdot T_\text{rev}) \times  \mathbf{U}_\text{ring}^\text{2\,sol}  \left(\vec c, T_\text{rev},\chi_\text{rot}^{\text{S}_1}, \chi_\text{rot}^{\text{S}_2} \right) \right]}_{\text{turn 2}} \times  
   \underbrace{\left[ \mathbf{U}_\text{WF} (t= T_\text{rev}) \times  \mathbf{U}_\text{ring}^\text{2\,sol}  \left(\vec c, T_\text{rev},\chi_\text{rot}^{\text{S}_1}, \chi_\text{rot}^{\text{S}_2}\right) \right]}_{\text{turn 1}} \times  \vec S_0\,.
 \end{split}
\label{eq:polarization-evolution-with-WF-and-two-solenoids}
\end{equation} 
\end{widetext}

\subsection{Spin-rotation angle in a static solenoid}

In a solenoidal magnet with a field integral $\text{BDL} = \int B_\parallel \dd \ell$, the spins are rotated around the longitudinal direction $\vec e_z$, and the rotation angle is given by 
\begin{equation}
\chi_\text{rot}^{\text{Sol}}  = - \frac{q}{m} \cdot \frac{(1 + G)}{\gamma \beta c} \int B_\parallel \dd \ell \,. 
\end{equation}
The spin rotation angle in the solenoid for deuterons at a momentum of $P=\SI{970}{MeV \per c}$, normalized to the magnetic field integral, amounts to 
\begin{equation}
\frac{\chi_\text{rot}^{\text{Sol}}}{\int B_\parallel \dd \ell} = \SI{-0.264872}{\radian \per \tesla \per \meter}\,.
\end{equation}

\subsection{Spin tune and spin closed orbit with solenoids using $\vec S_3(t)$}
\begin{figure*}[t]
\centering
\includegraphics[width=\textwidth]{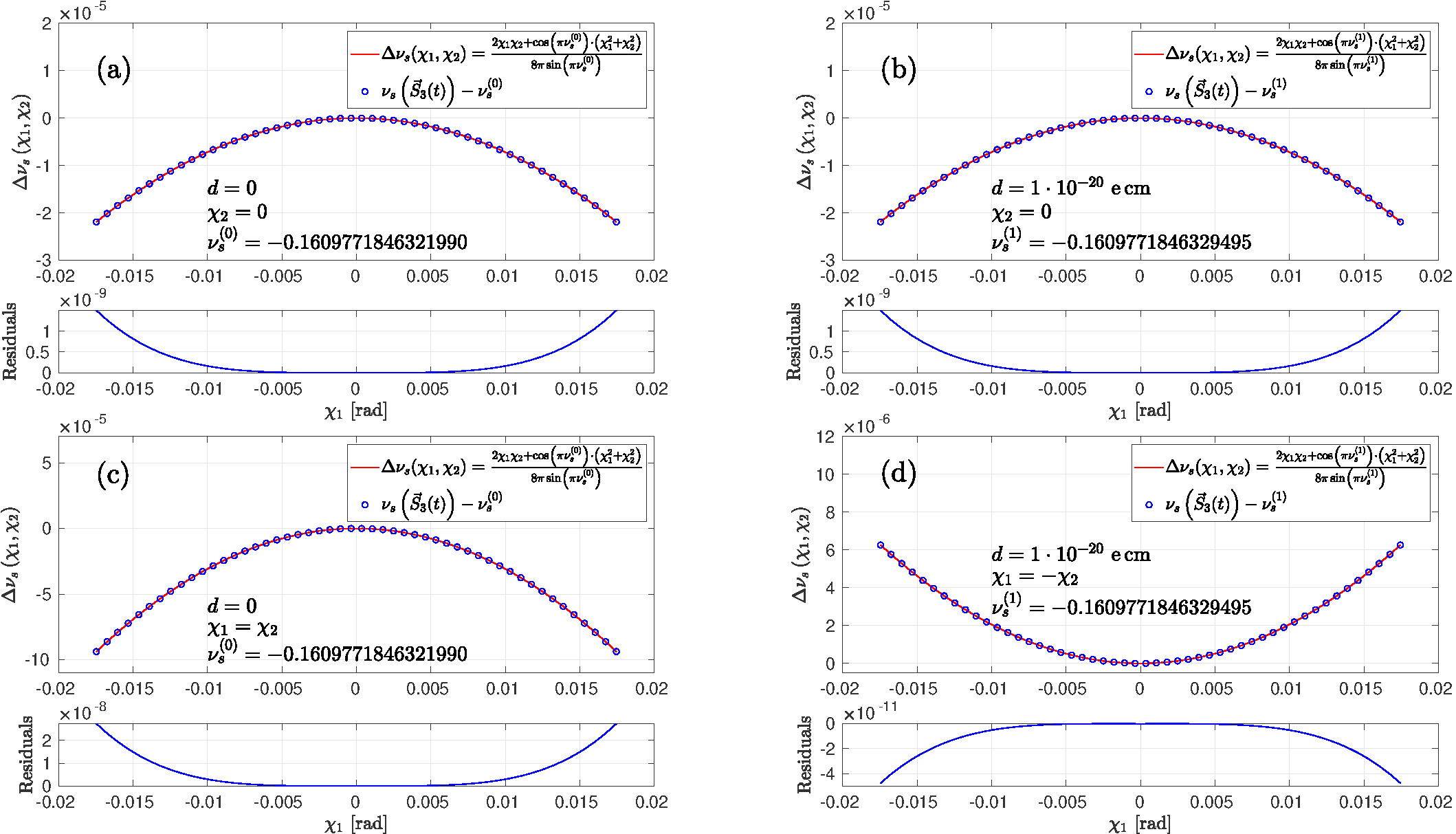}           
\caption{\label{fig:spin-tune-change-as-fct-of-chi1-and-chi2}
\small Change of the spin tune $\Delta \nu_s(\chi_1, \chi_2)$ for deuterons using solenoids in the machine (see Fig.\,\ref{fig:ring-with-two-more-solenoids}) under the conditions of Table\,\ref{tab:list-of-parameters} using Eq.\,(\ref{eq:spin-tune-calculated-from-S8}) and $\vec S_3(t)$ from Eq.\,(\ref{eq:polarization-evolution-with-WF-and-two-solenoids}). Panels (a) and (c) show for $d=0$  $\Delta \nu_s(\chi_1, \chi_2) = \nu_s(t) - \nu_s^{(0)}$, while (b) and (d) show for $d=\SI{e-20}{e.cm}$  $\Delta \nu_s(\chi_1, \chi_2) = \nu_s(t) - \nu_s^{(1)}$.
Panel (a) and (b): $\chi_2=0$,  c): $\chi_1=\chi_2$, and (d): $\chi_1 = -\chi_2$. $\nu_s^{(0)}$ and $\nu_s^{(1)}$ are given in the inserts [see also Eq.\,(\ref{eq:spin-tunes-numerical-values})]. Residuals show the difference between the simulations (\textcolor{blue}{$\circ$}) and the approximations  from Eq.\,(\ref{eq:spin-tune-change-as-fct-of-chi1-and-chi2}) (red lines).
}
\end{figure*}

In the following, the abbreviation, \textit{e.g.},  $\chi_\text{rot}^{\text{Sol 1}} = \chi_1$ is used. For an ideal ring, free of magnetic imperfections, the spin tune change $\Delta \nu_s (\chi_1, \chi_2)$, due to solenoids S$_1$ and S$_2$ in the ring (see Fig.\,\ref{fig:ring-with-two-more-solenoids}), the left side of Eq.\,(30) of Ref.\,\cite{PhysRevAccelBeams.20.072801} can be approximated by $\pi \Delta \nu_s(\chi_1, \chi_2) \cdot \sin(\pi \nu_s^0)$, where $\nu_s^0$ denotes the unperturbed spin tune in the machine. For small spin rotation angles in the solenoids, Eq.\,(30) can thus be approximated by
\begin{equation}
 \Delta \nu_s(\chi_1, \chi_2) = \frac{2\chi_1\chi_2 + \cos\left( \pi \nu_s^0 \right) \cdot \left( \chi_1^2 + \chi_2^2 \right) }{8\pi\sin \left(\pi\nu_s^0 \right) }\,.
 \label{eq:spin-tune-change-as-fct-of-chi1-and-chi2}
\end{equation}

In order to validate the spin evolution equation for $\vec S_3(t)$, given in Eq.\,(\ref{eq:polarization-evolution-with-WF-and-two-solenoids}), in Fig.\,\ref{fig:spin-tune-change-as-fct-of-chi1-and-chi2} the  spin tune changes $\Delta \nu_s$ are compared to the approximation of Eq.\,(\ref{eq:spin-tune-change-as-fct-of-chi1-and-chi2}) for four different cases.

\subsection{Spin-closed orbit in a non-ideal lattice}
The static solenoids or magnetic imperfections in the ring affect the spin-closed orbit vector $\vec n_s = \vec c$\, in the machine. The situation is similar to the one depicted in Fig.\,\ref{fig:tilt-angle-xi}, but there, only the tilt due to the EDM was  taken into account. The presence of static solenoids in the ring can be numerically evaluated using Eq.\,(\ref{eq:spin-closed-orbit-vector-as-fct-of-time}) with $\vec S_3(t)$ [Eq.\,(\ref{eq:polarization-evolution-with-WF-and-two-solenoids})]. 

Since the time $t$ begins to count right behind the RF Wien filter (see Fig.\,\ref{fig:ring-with-two-more-solenoids}), evaluation of Eq.\,(\ref{eq:spin-closed-orbit-vector-as-fct-of-time}) at $t = T_\text{rev}$ (or integer multiples of $T_\text{rev}$ [see Eq.\,(\ref{eq:mod-condition-to-evaluate-wf-once-per-turn})]), yields the orientation of the spin-closed orbit vector $\vec c$ at the RF Wien filter
\begin{equation}
 \vec c = \vec n_s(t=T_\text{rev})\,.
\end{equation}

Figure\,\ref{fig:spinclosedorbitatWF} shows how the axis $\vec c = (c_x, c_y, c_z)$ is affected by the solenoids S$_1$ and S$_2$, and the presence of an EDM $d$. 
\begin{figure*}[htb]
\centering
\includegraphics[width=\textwidth]{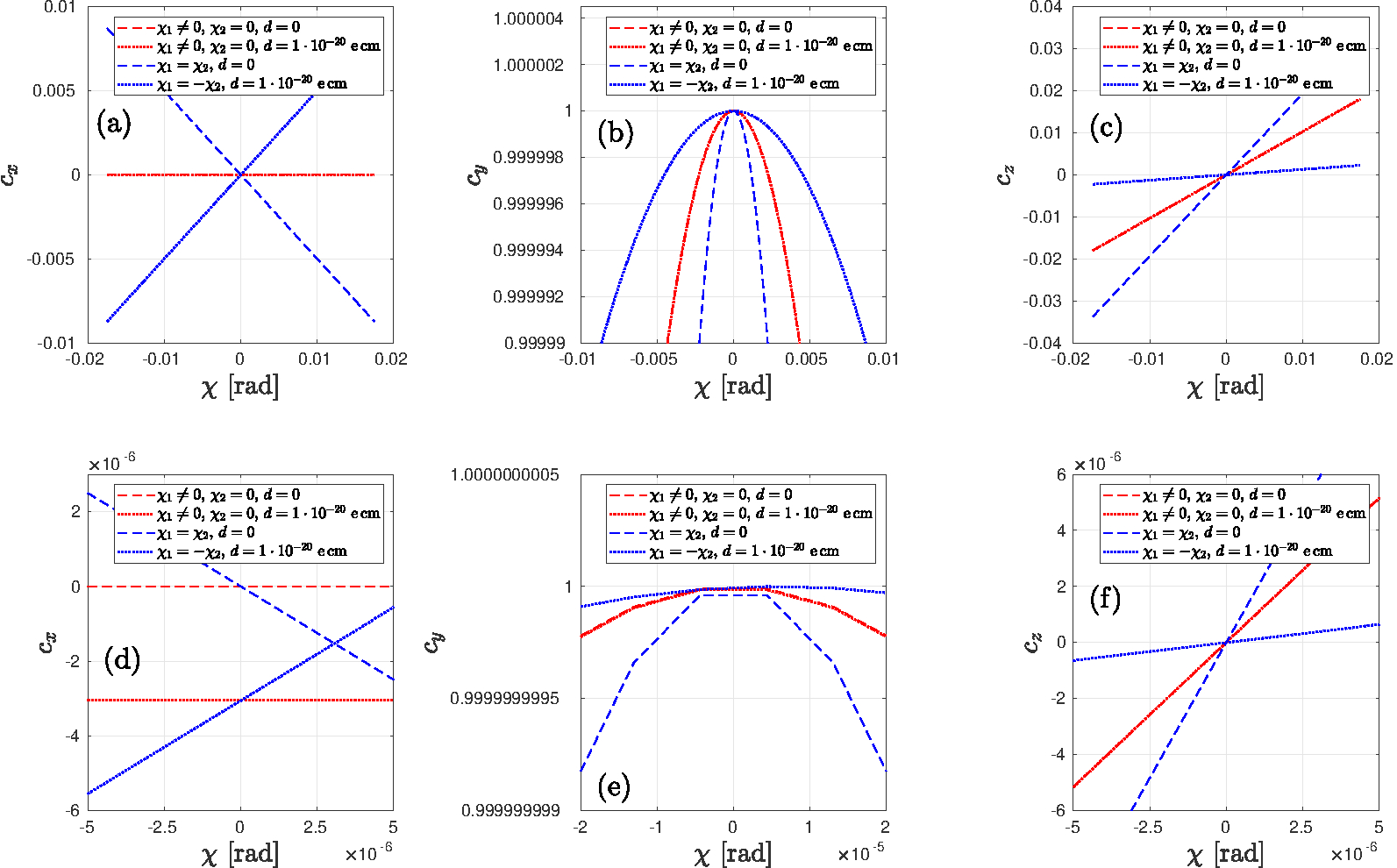}           
\caption{\label{fig:spinclosedorbitatWF} Six panels showing the components of $\vec c = (c_x, c_y, c_z)$ for different combinations of rotations in the solenoids, for deuterons at a momentum of $P=\SI{970}{MeV \per c}$.
}
\end{figure*}
For numerical comparisons, a number of special cases are numerically evaluated in Table\,\ref{table:various-chis-and-SCO-vector}.
\begin{table*}[htb]
\begin{ruledtabular}
\caption{ \label{table:various-chis-and-SCO-vector}
\small Components of the spin closed orbit vector $\vec c = (c_x, c_y, c_z)$ right at the RF Wien filter, for different settings of the solenoids S$_1$ and S$_2$ in the machine (see Fig.\,\ref{fig:ring-with-two-more-solenoids}).}
 \begin{tabular}{rrrrrr}
 $\chi_1$\,[\si{\degree}] & $\chi_2$\,[\si{\degree}] & $d$\,[e\, cm] & $c_x$                & $c_y$              & $c_z$              \\\hline
 $0$                        & $0$                        & 0           & \num{0.000000e+00}   & \num{1.000000e+00} & \num{0.000000e+00} \\
 $0$                        & $0$                        & \num{e-20}  & \num{-3.053662e-06}  & \num{1.000000e+00} & \num{4.255557e-17} \\     
 $1$                        & $0$                        & \num{e-20}  & \num{-3.053167e-06}  & \num{9.998378e-01} & \num{1.801136e-02} \\     
 $0$                        & $1$                        & \num{e-20}  & \num{-8.728505e-03}  & \num{9.998378e-01} & \num{1.575676e-02} \\     
 $1$                        & $1$                        & \num{e-20}  & \num{-8.724615e-03}  & \num{9.993921e-01} & \num{3.375307e-02} \\     
 $1$                        & $-1$                       & \num{e-20}  & \num{8.723460e-03}   & \num{9.999594e-01} & \num{2.254871e-03} \\     
 $-1$                       & $1$                        & \num{e-20}  & \num{-8.729567e-03}  & \num{9.999594e-01} & \num{-2.254871e-03} \\     
 $-1$                       & $-1$                       & \num{e-20}  & \num{8.718511e-03}   & \num{9.993922e-01} & \num{-3.375307e-02} 
\end{tabular}
\end{ruledtabular}
\end{table*}

\begin{figure*}[t]
\centering
\subfigure[\label{fig:ResonanceStrength-3by3a} 
$\left(\phi_\text{rot}^\text{WF}, \chi_\text{rot}^\text{Sol\,1}\right) = \left(\SI{-1}{\degree}, \SI{-1}{\degree}\right)$] 
  {\includegraphics[width=0.47\textwidth]{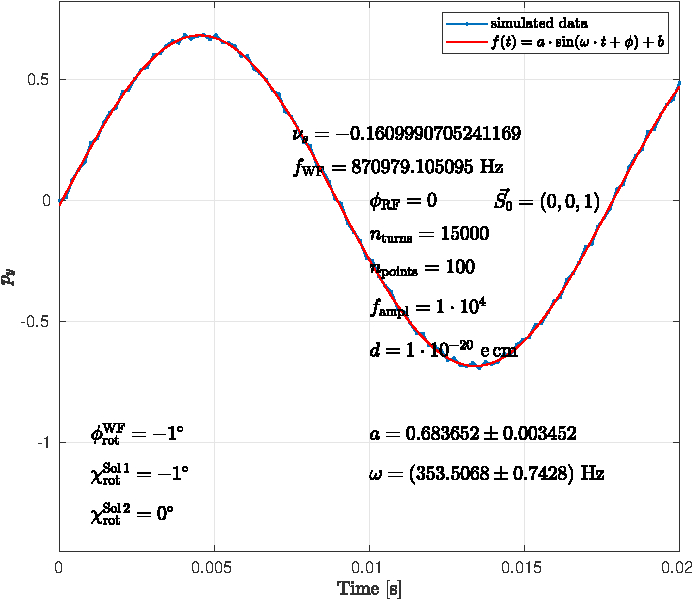}}
  \hspace{0.2cm}
\subfigure[\label{fig:ResonanceStrength-3by3e}  
  $\left(\phi_\text{rot}^\text{WF}, \chi_\text{rot}^\text{Sol\,1}\right) = \left(\SI{0}{\degree}, \SI{0}{\degree}\right)$]
  {\includegraphics[width=0.47\textwidth]{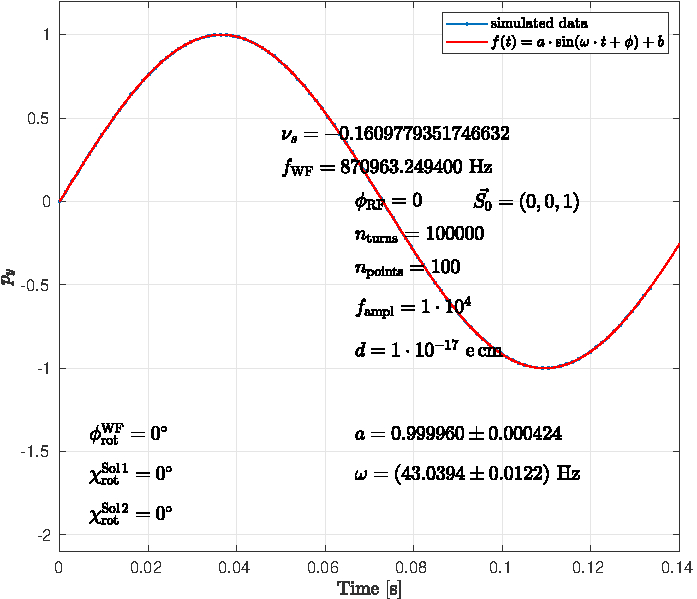}}
\caption{
\label{fig:ResonanceStrength-3by3} 
Two examples for the evolution of $p_y(t)$ using $\vec S_3(t)$ from Eq.\,(\ref{eq:polarization-evolution-with-WF-and-two-solenoids}) for different combinations of Wien filter and solenoid spin rotation angle, denoted by  
$\left(\phi_\text{rot}^\text{WF},\chi_\text{rot}^\text{Sol\,1}\right)$, where $\chi_\text{rot}^\text{Sol\,2} = 0$. The parameters used for the calculation are indicated in each panel. For the beam, the conditions of Table\,\ref{tab:list-of-parameters} apply. The Wien filter is operated at harmonic $K = -1$. The EDM assumed in panel (b) is $1000$ times larger than in (a). The ratio of the fitted oscillation amplitudes in panels (a) and (b) is compatible with the expectation of a factor $\sqrt{2}/2$ [see Eq.\,(\ref{eq:amplitude-ratio-sqrt2-over-2})]. 
}
\end{figure*}

\subsection{Strength of the EDM resonance}
As depicted in Fig.\,\ref{fig:spin-tunes}, and already discussed in Sec.\,\ref{sec:determination-of-spintune}, the operation of the RF Wien filter modulates the spin tune. While the \textit{average} spin tune is equal to the one obtained when the RF Wien filter is switched off, solenoids and magnet misalignments in the ring, however, affect the spin tune. Therefore, the spin-precession frequency and thus the frequency at which the RF Wien filter should be operated, differs from the unperturbed spin tune. The spin tune $\nu_s$ must be determined anew for every solenoid setting to ensure that the resonance frequency for the RF Wien filter  is given by  
\begin{equation}
 f_\text{WF} = \left( K + \nu_s  \right) \cdot f_\text{rev}\,, K \in \mathbb{Z} \,,
 \label{eq:WF-frequencies2}
\end{equation}
and this frequency needs to be used in $\psi(t))$ [Eq.\,(\ref{eq:psi-of-t-in-wien-filter-including-phase})], as it controls the RF Wien filter spin-rotation matrix $\mathbf{R}(\vec n_\text{WF}, \psi(t))$ [Eq.\,(\ref{eq:Wien-filter-matrix})]. 

The EDM resonance strength $\varepsilon^\text{EDM}$, actually a \textit{resonance tune}, is defined as the ratio of the angular frequency of the vertical polarization oscillation $\Omega^{p_y}$ induced by the EDM relative to the orbital angular frequency $\Omega^\text{rev}$,
\begin{equation}
 \varepsilon^\text{EDM} = \frac{\Omega^{p_y}}{\Omega^\text{rev}}\,.
 \label{eq:resonance-strength}
\end{equation}
Since  $\Omega_{p_y}$ corresponds to $\omega$ [first line in Eq.\,(\ref{eq:initial-slope-from-full-oscillation})], the resonance strength can in principle be determined from a single observation of $\Omega_{p_y}$.
Alternatively, the resonance strength can be determined from the last line in Eq.\,(\ref{eq:initial-slope-from-full-oscillation}) via
\begin{equation}
  \varepsilon^\text{EDM} = \frac{\left. \dot p_y(t) \right|_{t=0}}{a\,\cos\phi} \cdot \frac{1}{\Omega^\text{rev}} \,,
\label{eq:resonance-strength-from-phi-variation}
\end{equation}
but this requires that the initial slopes need to be determined as function of, \textit{e.g.}, $\phi = \phi_\text{RF}$. The statistical aspects of this will be further elucidated in Sec.\,\ref{sec:comparison-of-different-epslion-extractions}.

\subsubsection{Evolution of $p_y(t)$ as function of $\phi_\text{rot}^\text{\rm WF}$ and $\chi_\text{rot}^\text{\rm Sol\,1}$}
\label{sec:Evolution-of-py-as-function-of-phiWF-and-chiSol}
The EDM resonance strength $\varepsilon^\text{EDM}$ [Eq.\,(\ref{eq:resonance-strength})] manifests itself in the oscillation frequency, as illustrated in Fig.\,\ref{fig:ResonanceStrength-3by3} for two pairs of Wien filter rotation angle and  spin-rotation angle in solenoid S$_1$, $(\phi_\text{rot}^\text{WF},\chi_\text{rot}^\text{Sol\,1})$, where $\chi_\text{rot}^\text{Sol\,2} = 0$. 

The resulting oscillation pattern of $p_y$ is fitted using
\begin{equation}
f(t) = a \sin (\omega\,t + \phi) +b \,, 
\label{eq:function-fitted-to-oscillation-pattern-to-get-resonance-strength}
\end{equation}
amplitude $a$ and frequency $\omega$ are given in each panel, together with various other parameters. The calculation for the ideal ring situation in panel (b) uses a $\num{1000}$ times larger assumed EDM value of $d = \SI{e-17}{e.cm}$ and a larger number of turns $n_\text{turns} = \num{100000}$, in order to make the oscillations of $p_y(t)$ visible as well.


\subsubsection{Comparison of $\varepsilon^\text{\rm EDM}$ from $\Omega^{p_y}$ and $\dot p_y(t)|_{t=0}$ by variation of $\phi_\text{\rm RF}$}
\label{sec:comparison-of-different-epslion-extractions}
One would expect that the variation of the RF phase $\phi_\text{RF}$ will affect the resulting oscillation amplitudes $a$ and offsets $b$ of Fig.\,\ref{fig:ResonanceStrength-3by3}, while the oscillation frequencies $\omega$, and thus the resonance strengths $\varepsilon^\text{EDM}$ remain unchanged.

In the panels of Fig.\,\ref{fig:ResonanceStrength-3by3-phiRF-variation}, for the same combinations of $\left( \phi_\text{rot}^\text{WF}, \chi_\text{rot}^\text{Sol\,1} \right)$, shown in Fig.\,\ref{fig:ResonanceStrength-3by3}, $\dot p_y(t)|_{t=0}$ and the oscillation frequency $\omega$ are computed for 36 randomly picked values of $\phi_\text{RF}$. The graph illustrates that in the presence of solenoid fields and RF Wien filter misalignments, the determination of $\dot p_y(t)|{_{t=0}}$ by variation of $\phi_\text{RF}$, making use of Eq.\,(\ref{eq:resonance-strength-from-phi-variation}) yields results comparable to the direct determination of the resonance strength from the oscillation frequency $\Omega^{p_y}$ via Eq.\,(\ref{eq:resonance-strength}). The oscillation amplitudes $a$ and $\dot{p}_y|_{t=0}$ exhibit an identical dependence on $\phi^\text{RF}$, while the obtained resonant tune $\varepsilon^\text{EDM}$ remains constant over the whole range of $\phi^\text{RF}$.   
\begin{figure*}[htb]
\centering
\subfigure[\label{fig:ResonanceStrength-3by3-phiRF-variation-a}  
  $(\phi_\text{rot}^\text{WF}, \chi_\text{rot}^\text{Sol\,1}) = (\SI{-1}{\degree}, \SI{-1}{\degree})$] 
  {\includegraphics[width=0.49\textwidth]{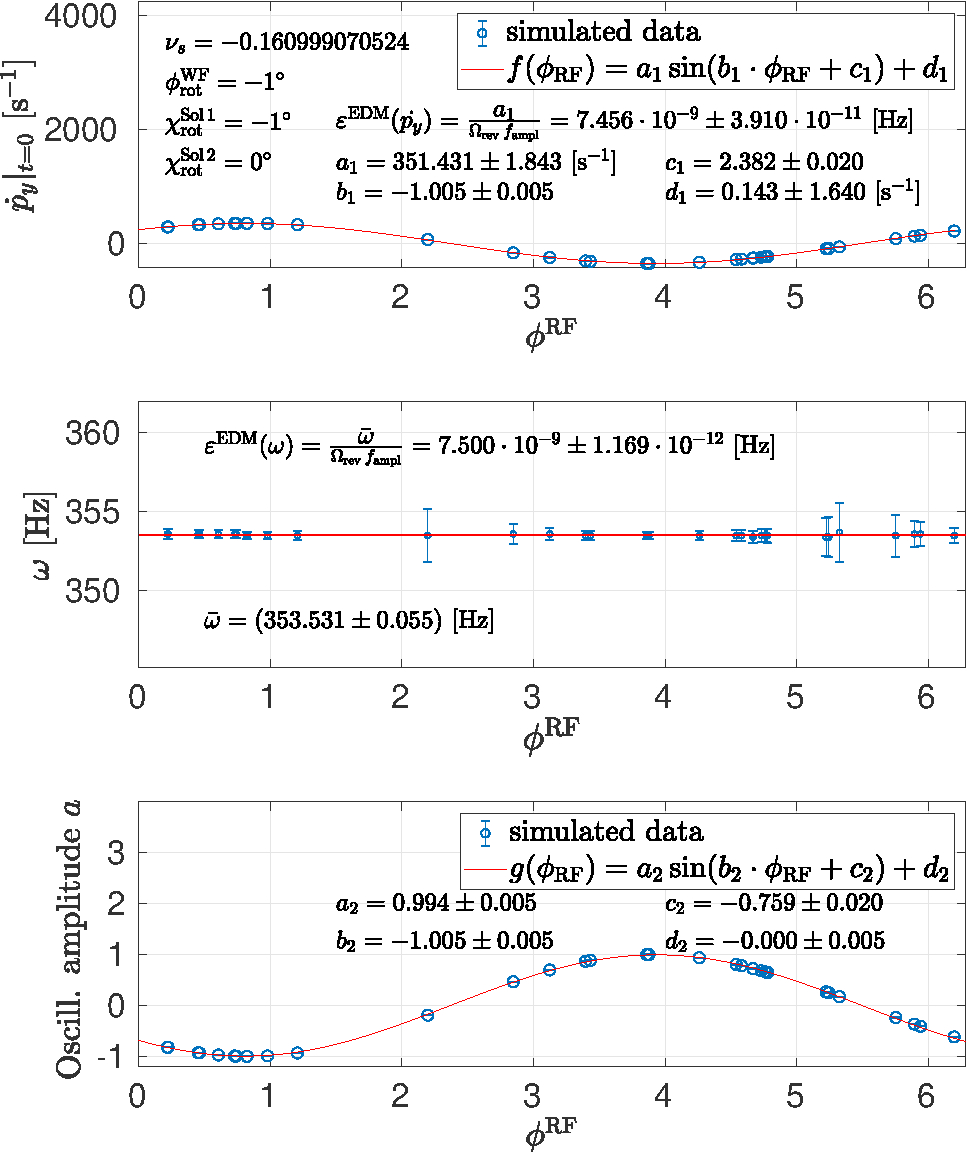}}
  \hspace{0.1cm}
\subfigure[\label{fig:ResonanceStrength-3by3-phiRF-variation-e}  
  $(\phi_\text{rot}^\text{WF}, \chi_\text{rot}^\text{Sol\,1}) = (\SI{0}{\degree}, \SI{0}{\degree})$, $n_\text{turns} = \num{e5}$,  $d = \SI{e-17}{e.cm}$.]
  {\includegraphics[width=0.49\textwidth]{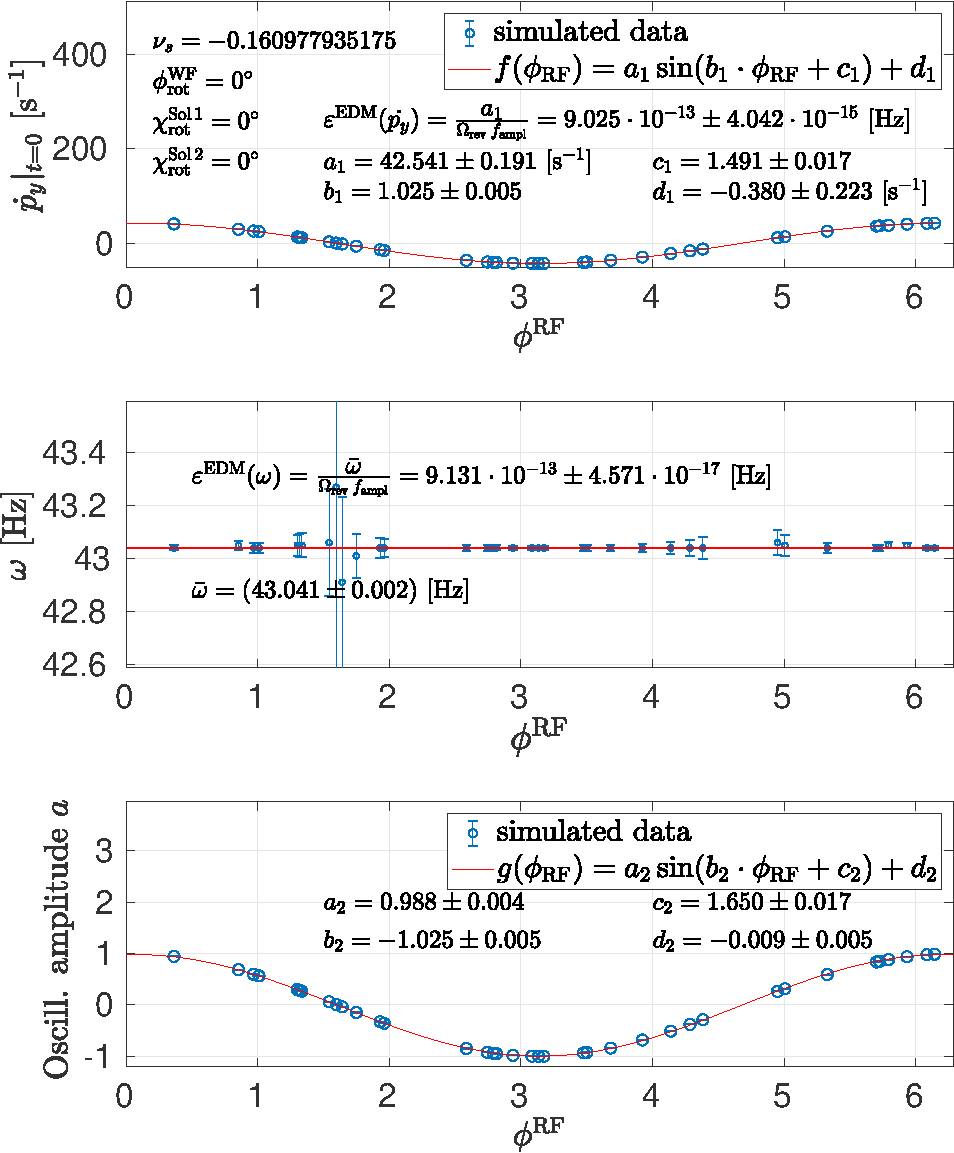}}
\caption{ \label{fig:ResonanceStrength-3by3-phiRF-variation}  Two examples showing 36 random values of $\phi_\text{RF}$ that are used to obtain the resonance strengths $\varepsilon^\text{EDM}$ from graphs like those shown in Fig.\,\ref{fig:ResonanceStrength-3by3} using Eqs.\,(\ref{eq:resonance-strength}) and (\ref{eq:resonance-strength-from-phi-variation}) for  combinations of the Wien filter and  solenoid spin rotation angle, denoted by $\left(\phi_\text{rot}^\text{WF},\chi_\text{rot}^\text{Sol\,1}\right)$. Depicted here as function of the randomly chosen $\phi_\text{RF}$  are the extracted initial slopes $\dot p_y(t)|_{t=0}$, $\omega = \Omega^{p_y}$, and the amplitude $a$ of the $p_y$ oscillation [Eq.\,(\ref{eq:function-fitted-to-oscillation-pattern-to-get-resonance-strength})]. The parameters used for the calculation are $n_\text{turns} = \num{2e4}$, $n_\text{points} = 200$, and $d = \SI{e-20}{e.cm}$. In panel (b), $n_\text{turns} = \num{e5}$, and the assumed EDM is $d = \SI{e-17}{e.cm}$ , \textit{i.e.}, $1000$ times larger than in (a), in order to enhance the effect. For the beam, the conditions of Table\,\ref{tab:list-of-parameters} apply. The RF Wien filter is operated at harmonic $K = -1$. The extracted resonance strengths are summarized in Table\,\ref{tab:comparison-Omega_y-todot-py}.}
\end{figure*}

The  resonance strengths extracted from $\dot p_y(t)|_{t=0}$ and $\Omega^{p_y}$ make use of the very same simulated data. The results are summarized in Table\,\ref{tab:comparison-Omega_y-todot-py}, where for the numbers that should match, the same color is used. Although the different extraction methods show good overall agreement,  the uncertainties of $\varepsilon^\text{EDM}(\Omega^{p_y})$, however, are substantially smaller than those from $\varepsilon^\text{EDM}(\dot p_y|_{t=0})$ by a factor of at least $20$. The reason for this is that in general frequencies can be measured more accurately than other quantities, and the determination of $\varepsilon^\text{EDM}(\Omega^{p_y})$ involves fewer uncertainties in the error propagation. The most accurate determinations  are obtained from $\Omega^{p_y}$ when $\chi_\text{rot}^\text{Sol\,1} = 0$. 
\begin{table*}[!]
\caption{\label{tab:comparison-Omega_y-todot-py} Resonance strengths extracted from Fig.\,\ref{fig:ResonanceStrength-3by3-phiRF-variation} for nine different combinations $\left( \phi_\text{rot}^\text{WF}, \chi_\text{rot}^\text{Sol\,1} \right)$ for an otherwise ideal COSY ring assuming a deuteron EDM of $d = \SI{e-20}{e.cm}$ (for (b), at $(\SI{0}{\degree}, \SI{0}{\degree})$, $d = \SI{e-17}{e.cm}$). The beam conditions are given in Table\,\ref{tab:list-of-parameters} using the \textit{real} field magnitudes of the RF Wien filter, since $f_\text{ampl}$ has been divided out. For the calculations $n_\text{turns} = \num{2e4}$ and $n_\text{points} = 200$, except for (b) , where $n_\text{turns} = \num{e5}$.} 
\renewcommand{\arraystretch}{1.25}
\begin{ruledtabular} 
 \begin{tabular}{rlccc}
\multicolumn{2}{c}{[$\num{e-11}$ Hz]} & \multicolumn{3}{c}{ $\left(\phi_\text{rot}^\text{WF},\chi_\text{rot}^\text{Sol\,1}\right)$}\\\hline
&                                    & $(-\SI{1}{\degree}, -\SI{1}{\degree})$ & $\left( \SI{0}{\degree}, -\SI{1}{\degree} \right)$      & $(\SI{1}{\degree}, -\SI{1}{\degree})$ \\
\multirow{ 2}{*}{$\varepsilon^\text{EDM}$} & from $\dot p_y|_{t=0}$ & \textcolor{blue}{$\num{745.563 \pm 3.910}$} & \textcolor{magenta}{$\num{539.778 \pm 1.695}$}  & \textcolor{blue}{$\num{750.455 \pm 3.312}$}  \\
                                           & from $\Omega^{p_y}$    & \textcolor{blue}{$\num{750.017 \pm 0.117}$} & \textcolor{magenta}{$\num{538.659 \pm 0.099}$}  & \textcolor{blue}{$\num{749.840 \pm 0.128}$}  \\\hline
&                                    & $(-\SI{1}{\degree}, \SI{0}{\degree})$  & $(\SI{0}{\degree}, \SI{0}{\degree})$       & $(\SI{1}{\degree}, \SI{0}{\degree})$ \\
\multirow{ 2}{*}{$\varepsilon^\text{EDM}$} & from $\dot p_y|_{t=0}$ & \textcolor{red}{$\num{517.167 \pm 2.741}$} & $(\num{90.251 \pm 0.404})\cdot \num{e-3}$  & \textcolor{red}{$\num{518.440 \pm 2.284}$} \\     
                                           & from $\Omega^{p_y}$    & \textcolor{red}{$\num{521.890 \pm 0.001}$} & $(\num{91.312 \pm 0.005})\cdot \num{e-3}$  & \textcolor{red}{$\num{521.681 \pm 0.001}$} \\\hline
&                                   & $(-\SI{1}{\degree}, \SI{1}{\degree})$  & $(\SI{0}{\degree}, \SI{1}{\degree})$       & $(\SI{1}{\degree}, \SI{1}{\degree})$ \\
\multirow{ 2}{*}{$\varepsilon^\text{EDM}$} & from $\dot p_y|_{t=0}$ & \textcolor{blue}{$\num{748.511 \pm 3.249}$} & \textcolor{magenta}{$\num{540.799 \pm 3.136}$}  & \textcolor{blue}{$\num{749.413 \pm 3.891}$}\\
                                           & from $\Omega^{p_y}$    & \textcolor{blue}{$\num{749.960 \pm 0.121}$} & \textcolor{magenta}{$\num{538.619 \pm 0.129}$}  & \textcolor{blue}{$\num{749.842 \pm 0.113}$} 
\end{tabular}
\end{ruledtabular}
\end{table*}

In the following, we briefly comment on some features of the results obtained so far (Fig.\,\ref{fig:ResonanceStrength-3by3}, Table\,\ref{tab:comparison-Omega_y-todot-py}). We observe that numerically $2\sin \pi \nu_s = 1.0041 \simeq 1$. Then, according to Appendix\,\ref{sec:appendixA}, we expect
\begin{equation}
 a( \SI{-1}{\degree}, \SI{-1}{\degree}) = \cos\left(\frac{\pi}{4}\right) \cdot a( \SI{0}{\degree}, \SI{0}{\degree})\,,
 \label{eq:amplitude-ratio-sqrt2-over-2}
\end{equation}
in good agreement with the results shown in Fig.\,\ref{fig:ResonanceStrength-3by3-phiRF-variation}. The resonance tunes determined from $\dot{p}_y|_{t=0}$ and from $\Omega^{p_y}$ are identical. For the above reason of  $2\sin \pi \nu_s \simeq 1$ and small EDM contribution, the equalities
\begin{equation}
\begin{split}
 \varepsilon^\text{EDM}(\SI{-1}{\degree},\SI{-1}{\degree}) & = \varepsilon^\text{EDM}(\SI{1}{\degree},\SI{1}{\degree}) \,,\text{ and} \\
  \varepsilon^\text{EDM}(\SI{\pm 1}{\degree},\SI{-1}{\degree}) & = \sqrt{2} \cdot  \varepsilon^\text{EDM}(\SI{-1}{\degree},\SI{0}{\degree})
\end{split}
\label{eq:simple-relationships}
\end{equation}
hold. 

\subsection{Resonance strength $\varepsilon^\text{EDM}$ for random points $\left(\phi_\text{rot}^\text{WF},\chi_\text{rot}^\text{Sol\,1}\right)$}
\begin{figure*}[tb]
\centering
\subfigure[\label{fig:ResonanceStrengthc}  $\varepsilon^\text{EDM}$ for $d = \SI{e-18}{e.cm}$.]
{\includegraphics[width=0.95\columnwidth]{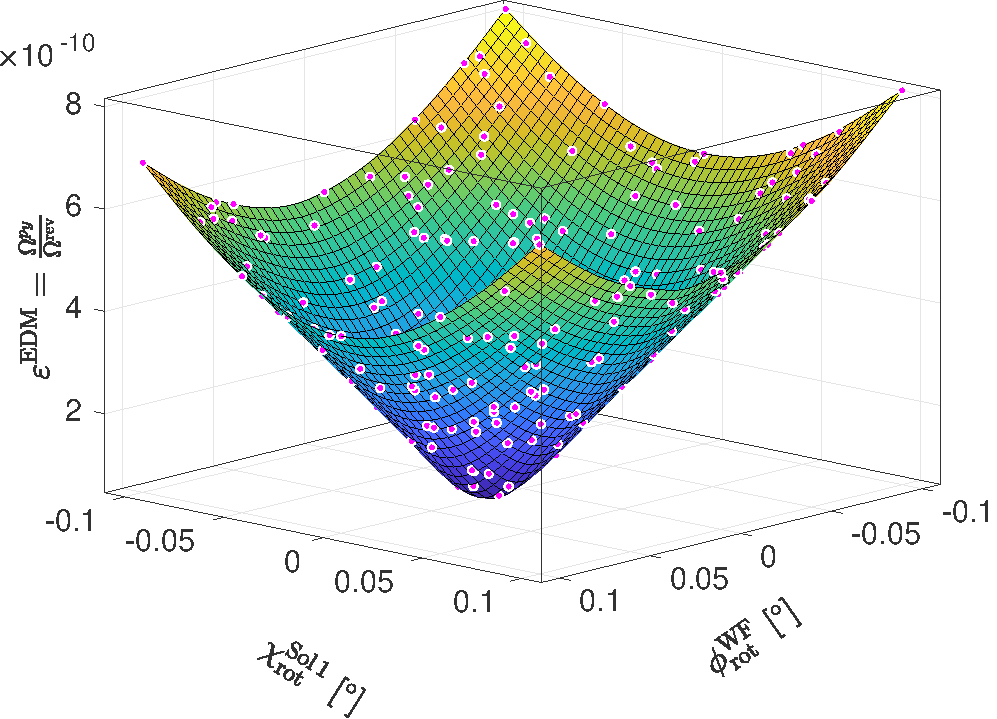}}
\hspace{0.3cm}
\subfigure[\label{fig:ResonanceStrengthd}   Contour plot of panel (a).]
{\includegraphics[width=0.95\columnwidth]{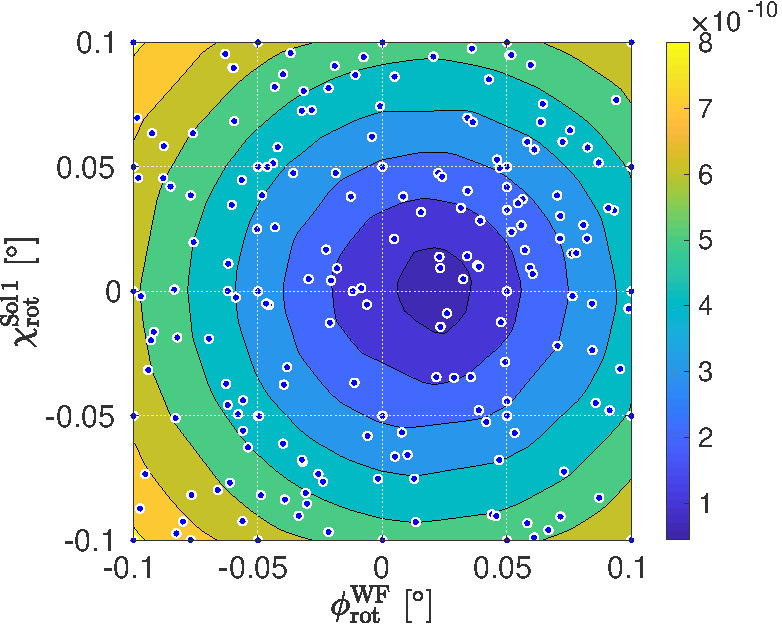}}
\caption{ \label{fig:ResonanceStrength}  
Panels (a) and (b) show the resonance strengths $\varepsilon^\text{EDM}$ on a grid in the range $\phi^\text{WF}_\text{rot} =  [-0.1\,\si{\degree}, \ldots, +0.n1\,\si{\degree}]$ and  $\chi^{\text{Sol}\, 1}_\text{rot} = [-0.1\,\si{\degree}, \ldots, +0.1\,\si{\degree}]$ with an  assumed EDM of $d = \SI{e-18}{e.cm}$. Each point in panels (a) and (b) is obtained from a calculation with 
 $n_\text{turns} = \num{200000}$ and $n_\text{points} = 100$. }
\end{figure*}

The resonance strengths shown in Fig.\,\ref{fig:ResonanceStrength} are obtained using the fit function of Eq.\,(\ref{eq:function-fitted-to-oscillation-pattern-to-get-resonance-strength}) ($\omega = \Omega^{p_y}$) and then Eq.\,(\ref{eq:resonance-strength}) for a set of randomly chosen pairs of $(\phi_\text{rot}^\text{WF},\chi_\text{rot}^\text{Sol\,1})$ and $\chi_\text{rot}^\text{Sol\,2} = 0$. For all points, $\phi_\text{RF}=0$ and $\vec S_0 = (0,0,1)$ 

Using the evolution function $\vec S_3(t)$ [Eq.\,\ref{eq:polarization-evolution-with-WF-and-two-solenoids}] which includes the ideal ring with solenoid S$_1$ and the RF Wien filter and an assumed EDM of $\SI{e-18}{e.cm}$, for which the EDM tilt angle is $\xi_\text{EDM} \approx  \SI{300}{\micro \radian}$, in the angular range, $\phi^\text{WF}_\text{rot} =  [-0.1\,\si{\degree}, \ldots, +0.1\,\si{\degree}]$,  $\chi^{\text{Sol}\, 1}_\text{rot} = [-0.1\,\si{\degree}, \ldots, +0.1\,\si{\degree}] $, and $\chi_\text{rot}^\text{Sol\,2} = 0$, the pattern shift is clearly visible, as seen in Fig.\,\ref{fig:ResonanceStrengthd}.

The relative uncertainties of the points shown in Fig.\,\ref{fig:ResonanceStrength} were obtained from the fits. In 
panels\,\ref{fig:ResonanceStrengthc} and \ref{fig:ResonanceStrengthd}, $\Delta \varepsilon^\text{EDM}/ \varepsilon^\text{EDM}$ ranges from \num{2.0e-05} to \num{4.1e-2}. 


For the set of points $\left(\phi_\text{rot}^\text{WF}, \chi_\text{rot}^\text{Sol\,1}\right)$ shown in Fig.\,\ref{fig:ResonanceStrength}, the initial spin tunes $\nu_s$, \textit{i.e.}, before the RF WF is turned on, are shown in Fig.\,\ref{fig:spintune}. The result indicates the familiar quadratic dependence $\Delta \nu_s(\chi_1, \chi_2 = 0) \propto \chi_1^2$, described by Eq.\,(\ref{eq:spin-tune-change-as-fct-of-chi1-and-chi2}). 
\begin{figure}[tb]
\centering
 \includegraphics[width=1\columnwidth]{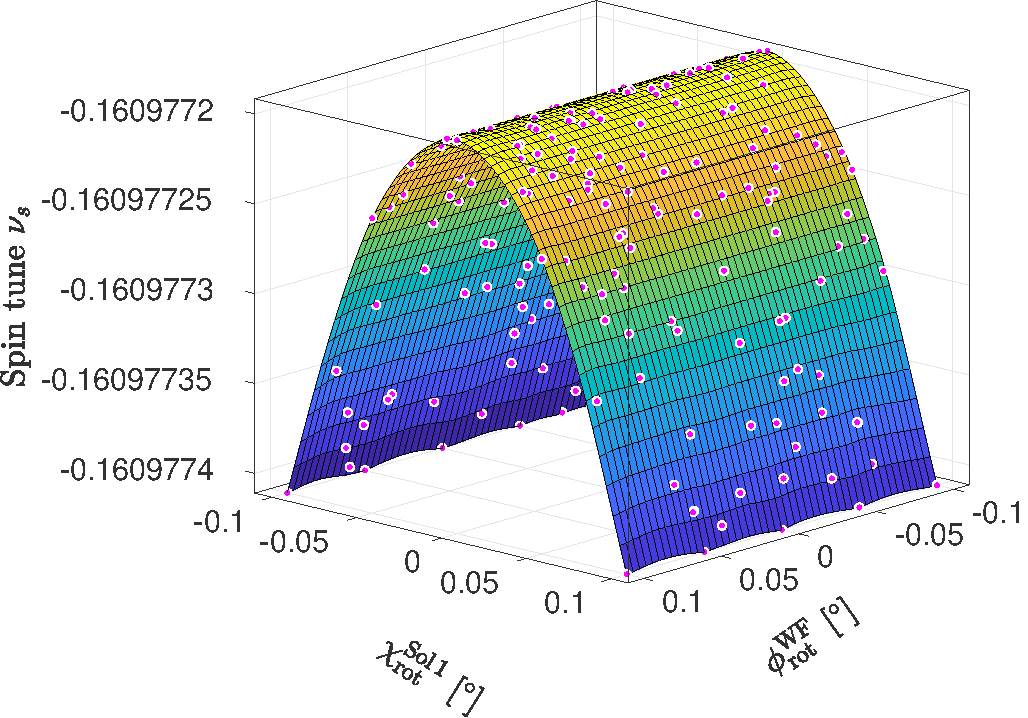}
\caption{ \label{fig:spintune} Initial spin tunes $\nu_s$ for the angular intervals $\phi^\text{WF}_\text{rot} = \chi^{\text{Sol}\, 1}_\text{rot} = \interval{ -0.1\,\si{\degree}} { \ldots, +0.1\,\si{\degree} }$ for the data points $\left( \phi^\text{WF}_\text{rot}, \chi^{\text{Sol}\, 1}_\text{rot} \right)$ shown in Figs.\,\ref{fig:ResonanceStrengthc} and \ref{fig:ResonanceStrengthd} with an assumed EDM of $d = \SI{e-18}{e.cm}$.}
\end{figure}
\subsection{Characterization of $\varepsilon^\text{EDM}\left(\phi_\text{rot}^\text{WF}, \chi_\text{rot}^\text{Sol\,1}\right)$ }
\subsubsection{Operation of RF Wien filter exactly on resonance}
In this section, the contour of the surface $\varepsilon^\text{EDM}\left(\phi_\text{rot}^\text{WF}, \chi_\text{rot}^\text{Sol\,1}\right)$, shown in Fig.\,\ref{fig:ResonanceStrengthc}, is compared to the theoretical expectation, given in Eq.\,(\ref{eq:EpsilonMap}). The functional dependence describes a quadratic surface, also know as \textit{Elliptic Paraboloid}, and is used here in the form
\begin{equation}
\begin{split}
 \left({\varepsilon^\text{EDM}}\right)^2 = \quad & A \cdot \left( \phi_\text{rot}^\text{WF} - \phi_0 \right)^2 \\
 &+  B\cdot \left(\frac{\chi_\text{rot}^\text{Sol\,1}}{2\sin\pi\nu_s^{(2)}} + \chi_0\right)^2 + C \,,
 \label{eq:surface-fit-function-for-epsilon}
\end{split} 
\end{equation}
where the unperturbed spin tune $\nu_s^{(2)}$ for the EDM of $d = \SI{e-18}{e.cm}$, assumed in the simulation, is given by
\begin{equation}
\begin{split}
 \nu_s^{(2)} & = \num{-0.160977192137641}\,, \text{and} \\
 2\sin\pi\nu_s^{(2)} & = -\num{0.968883216683076}\,.
 \end{split}
\end{equation}

It should be emphasized that the simulations shown in Fig.\,\ref{fig:ResonanceStrength} reflect the situation when the RF Wien filter is operated \textit{exactly on resonance}. During the corresponding EDM experiments in the ring, however, a certain spin-tune feedback is imperative to maintain for long periods of time the resonance condition, \textit{i.e.}, the spin-precession frequency in Eq.\,(\ref{eq:psi-of-t-in-wien-filter-including-phase}), using the measured spin tune\,\cite{PhysRevLett.115.094801}. To maintain phase \textit{and} frequency when the RF Wien filter is actively operating, turns out to be much more tricky, and more sophisticated approaches, beyond those outlined in\,\cite{PhysRevLett.119.014801}, are presently being pursued by the JEDI collaboration. Only such a phase \textit{and} frequency lock during a measurement cycle enables one to take full advantage of the large spin-coherence time (SCT) of $\tau_\text{SCT} \simeq \SI{1000}{s}$, achieved by JEDI at COSY\,\cite{Guidoboni:2016bdn,Guidoboni:2017ayl}.

The result of a fit without weighting is shown in Fig.\,\ref{fig:ResonanceStrengthc-FIT}. It should be noted that within the uncertainties obtained from the fit, $A=B$, while $C$ and $\chi_0$ are compatible with zero. Here, $\chi_0$ represents a primordial tilt of the stable spin axis at the RF Wien filter along the horizontal axis, $c_x$. For the model ring, one would expect
\begin{equation}
\chi_0 = 0 = c_x\,,
\end{equation}
a property which is nicely returned by the fit shown in Fig.\,\ref{fig:ResonanceStrengthc-FIT}. 
\begin{figure*}[htb]
 \centering
\subfigure[\label{fig:ResonanceStrengthc-FIT} Fit to the surface of $\left(\varepsilon^\text{EDM}\right)^2$, shown in Fig.\,\ref{fig:ResonanceStrengthc}, using Eq.\,(\ref{eq:surface-fit-function-for-epsilon}). The resonance strengths have been scaled by a factor around \num{6e9}.]
 {\includegraphics[width=\columnwidth]{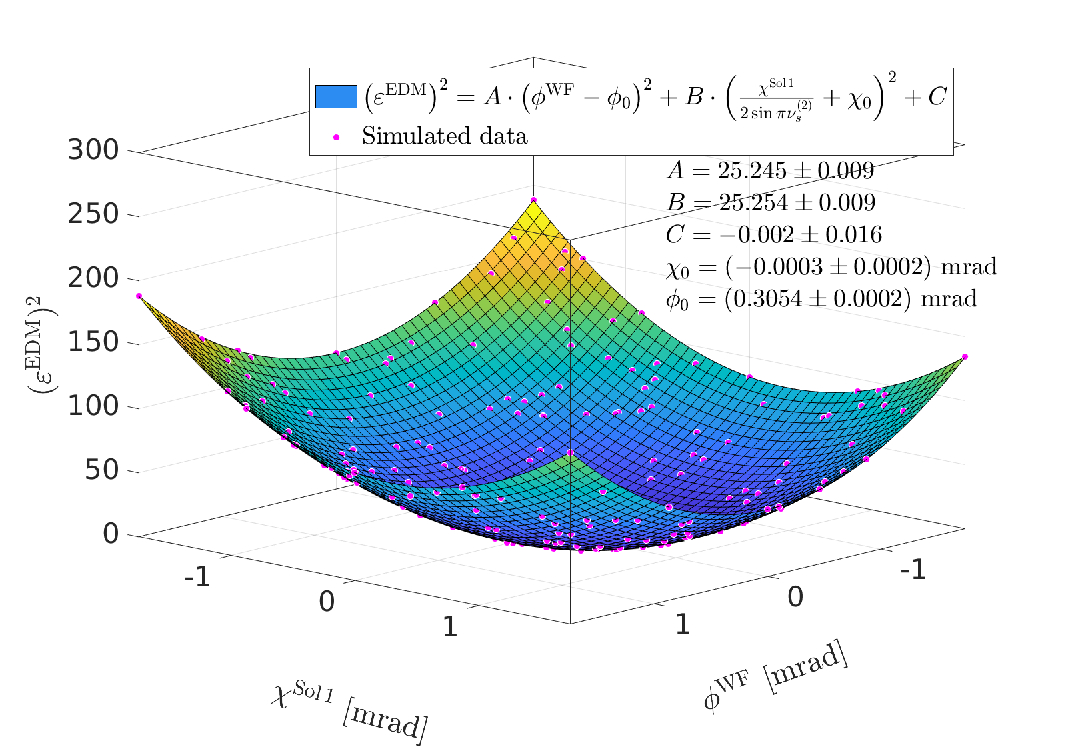}}
 \subfigure[\label{fig:ResonanceStrengthc-FIT2}  Fit to the simulated data from Fig.\,\ref{fig:ResonanceStrengthc}, using Eq.\,(\ref{eq:surface-fit2-function-for-epsilon}) with $F = \num{e20}$.]
 {\includegraphics[width=\columnwidth]{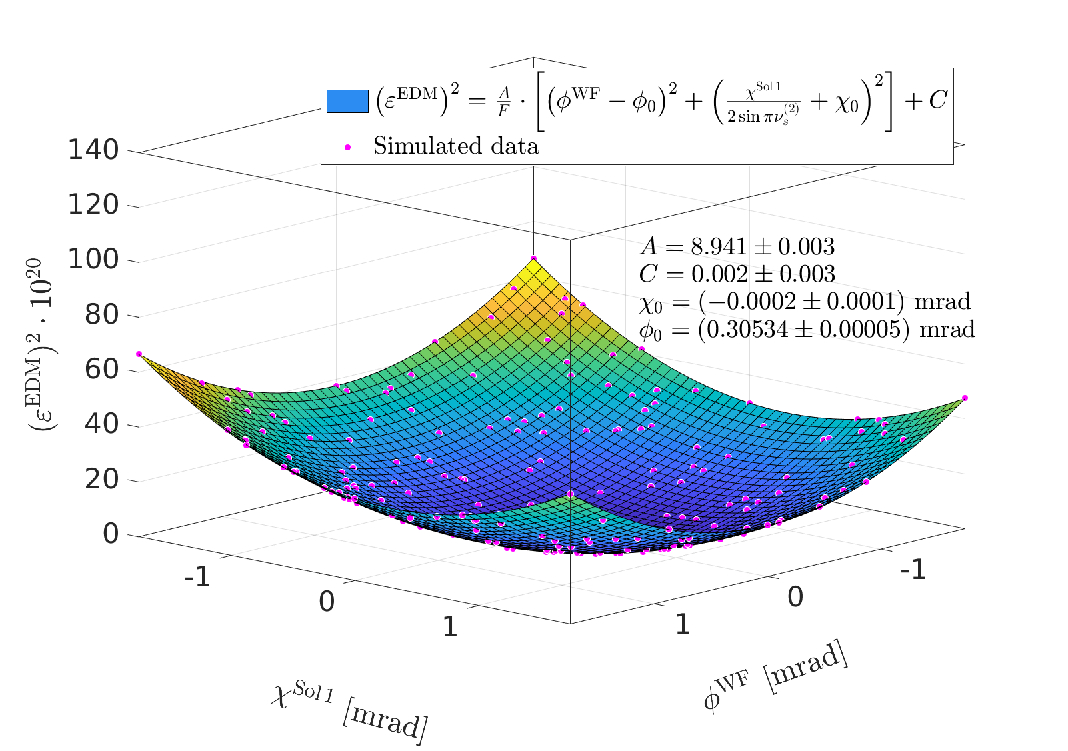}}
\caption{\label{fig:ResonanceStrengthc-FIT-and-FIT2} Fits to the simulated data for the resonance strength  $\left(\varepsilon^\text{EDM}\right)^2$ as function of $\left(\phi_\text{rot}^\text{WF}, \chi_\text{rot}^\text{Sol\,1}\right)$.}
\end{figure*}

In addition, the fit to the simulated data is expected to return $\phi_0 = \left|\xi_\text{EDM}(d = \SI{e-18}{e.cm})\right| =  \SI{0.3054}{\milli \radian}$, given by Eq.\,(\ref{eq:xiEDM}), and the fitted result 
\begin{equation}
 \phi_0 =  (\num{0.3054} \pm \num{0.0002}) \, \si{{\milli \radian}}
\end{equation}
returns this value accurately.

\subsubsection{Validation of the scale of $\varepsilon^\text{\rm EDM}$}

The fit with the elliptic paraboloid, shown in Fig.\,\ref{fig:ResonanceStrengthc-FIT}, indicates that the surface is described with $A = B$. In the following, the first fit function from Eq.\,(\ref{eq:surface-fit-function-for-epsilon}) is slightly altered, yielding
\begin{equation}
\begin{split}
 \left({\varepsilon^\text{EDM}}\right)^2  = \frac{A}{F}  \cdot  
\Biggl[ &  \left( \phi^\text{WF}_\text{rot} - \phi_0 \right)^2  \\
 &  +  \left(\frac{\chi_\text{rot}^\text{Sol\,1}}{2\sin\pi\nu_s^{(2)}} + \chi_0\right)^2  \Biggr] + C\,,
 \label{eq:surface-fit2-function-for-epsilon}
\end{split} 
\end{equation}
where a factor $F = \num{e20}$ has been used to scale the resonance strength.  The second fit now uses weights derived from the uncertainty of the fitted $\Omega^{p_y}$  using Eq.\,(\ref{eq:resonance-strength}). The resulting fit is shown in Fig.\,\ref{fig:ResonanceStrengthc-FIT2}. The agreement between theoretical model and simulated data is good, the $\chi^2/\text{ndf} = 374.4/194 = 1.9$.

According to Eq.\,(\ref{eq:EpsilonMap}), the factor in front of the brackets in Eq.\,(\ref{eq:surface-fit2-function-for-epsilon})  reads
\begin{equation}
\frac{A}{F} =  k \stackrel{!}{=} \frac{ \psi_\text{WF}^2}{16\pi^2} \,,
\end{equation}
where the Wien filter rotation angle $\psi_\text{WF}$ from Eq.\,(\ref{eq:psi-angle-in-WF}) is used. Inserting the numerical value of $A$ from the fit (inset Fig.\,\ref{fig:ResonanceStrengthc-FIT2}), and taking into account that the results are in \si{\milli \radian}, the ratio 
\begin{equation}
 \frac{A \cdot \num{e6}}{F \cdot k} = \num{9.9954e-01}
\end{equation}
yields the expected value near unity, which validates the scaling factor in Eq.\,(\ref{eq:EpsilonMap}). 

The second fit yields a similar value for
\begin{equation}
\begin{split}
  \phi_0  & =  (\num{0.30534} \pm \num{0.00005}) \, \si{\milli \radian} \\
                    & \approx \left|\xi_\text{EDM}(d = \SI{e-18}{e.cm})\right| \,,
\end{split}
\end{equation}
compared to the first fit, shown in Fig.\,\ref{fig:ResonanceStrengthc-FIT}, and  $\chi_0$ and $C$ are both compatible with zero.

\section{Conclusions and outlook}
\label{sec:conclusions}

The $\mathbf{SO(3)}$ matrix formalism used here to describe the spin rotations on the closed orbit, \textit{i.e.,} the spin dynamics of the interplay of an RF Wien filter with a machine lattice that includes solenoids, proved very valuable. The general features of the deuteron EDM experiment at COSY can be obtained rather immediately. Of course, the approach taken is no replacement for more advanced spin-tracking codes, but the results obtained here can be applied to benchmark those codes. 

In addition, it should be noted that the JEDI collaboration is presently applying beam-based alignment techniques to improve the knowledge about the absolute beam positions  in COSY\,\cite{TWagner}. Once this is accomplished, the approach described here to parametrize the spin rotations \textit{solely} on the basis of the closed orbit, will become more realistic. 

The polarization evolution in the ring in the presence of an RF Wien filter that is operated on resonance, in terms of the \textit{resonance tune} or \textit{resonance strength} $\varepsilon^\text{EDM}$ is theoretically well understood. This will allow us  to investigate in the future effects of increasingly smaller  magnetic imperfections, either through additional \textit{solenoidal} fields in the ring, or by \textit{transverse} magnetic fields via the rotation of the RF Wien filter around its axis.

In the near future, it is planned to incorporate into the developed matrix formalism also dipole magnet displacement and rotation parameters, available from a recent survey at COSY. This will allow us to determine the orientation of the stable spin axis of the machine at the location of the RF Wien filter, and to extract the EDM from a measurement of the resonance strengths as function of $\left(\phi_\text{rot}^\text{WF}, \chi_\text{rot}^\text{Sol\,1}\right)$. It is possible to incorporate the spin rotations from misplaced and rotated quadrupole magnets on the closed orbit into the formalism as well. 

An approach based on the polynomial chaos expansion has been successfully applied to determine a hierarchy of uncertainties during the construction of the RF Wien filter\,\cite{Slim:2017bic}. Such a methodology, in conjunction with the spin-tracking approach based on the matrix formalism outlined here, can be employed to efficiently generate a hierarchy of uncertainties for the EDM prototype ring\,\cite{1812.08535} from the different design parameters of the ring. 

The spin-tracking approach used here, shall be also applied to study various aspects of the presently applied spin-tune feedback system, used to phase-lock the spin precession to the RF of the Wien filter\,\cite{PhysRevLett.119.014801}. 

\section*{Acknowledgment}
This work has been performed in the framework of the JEDI collaboration and is supported by an ERC Advanced-Grant of the European Union (proposal number 694340). The work of N.N.N. was a part of the Russian MOS program 0033-2019-0005. Numerous discussions with colleagues went into this document, foremost we would like to acknowledge those with Volker Hejny, Alexander Nass, Jörg Pretz,  and Artem Saleev.

\begin{appendix}
\section{Dependence of the EDM resonance strength on $\phi^\text{WF}$ and $\chi^\text{Sol\,1}$}
\label{sec:appendixA}

The functional dependence of a physical rotation of the Wien filter around the beam axis by $\phi^\text{WF}_\text{rot}$ and of a spin rotation in static solenoids (see Fig.\,\ref{fig:sketch-ring-and-WF}) by $\chi_\text{rot}^{\text{Sol}_{1,2}}$ on the resonance strength $\varepsilon^\text{EDM}$ [Eq.\,(\ref{eq:resonance-strength})] is discussed.

At the location of the polarimeter, only the vertical and radial components of the beam polarization [$S_y(t)$  and $S_x(t)$]  can be determined. At the RF Wien filter, the orientation of the stable spin axis is denoted by $\vec c$, and in EDM mode the direction of the magnetic field by $\vec n_\text{WF}$  [see Eq.\,(\ref{eq:definition-of-EDM-mode})]. The in-plane $S_x(t)$ thus obviously depends on $[\vec n_\text{WF} \times \vec c\,]$. 

In an ideal all-magnetic ring under consideration, the stable spin axis is close to the vertical direction $\vec e_y$, 
\begin{equation}
\begin{split}
\vec{c} & =       \cos\xi_\text{EDM} \cdot \vec e_y + \sin\xi_\text{EDM} \cdot \vec e_x \\
        & \approx \vec e_y + \xi_\text{EDM} \cdot \vec e_x\,.
\end{split}
\end{equation}
In EDM mode, the magnetic axis of the RF Wien filter can be approximated by
\begin{equation}
\begin{split}
 \vec n_\text{WF} & =       \cos \phi^\text{WF}_\text{rot} \cdot \vec e_y + \sin \phi^\text{WF}_\text{rot} \cdot \vec e_x \\
                 & \approx \vec e_y + \phi^\text{WF}_\text{rot} \cdot \vec e_x \,.
\end{split}
\end{equation}
			
The stable spin axis $\vec c$ can be manipulated by static solenoids in the ring, and the drift solenoids $S_{1,2}$ of the electron coolers (or the Siberian snake instead of S$_1$) generate the spin kicks $\chi_{1,2}$. When both solenoids $S_{1,2}$ are turned on, one can write for the stable spin axis
\begin{equation}
	\begin{split}
		c_x & = \xi_\text{EDM} + \frac{1}{2}\chi_2\, , \\
		c_z & = \frac{1}{2\sin \pi \nu_s}( \chi_1 + \chi_2 \cos \pi \nu_s)\, .
	\end{split}
\end{equation}

In case solenoid S$_2$ is off ($\chi_2 = 0$), one obtains
\begin{equation}
 	[\vec n_\text{WF} \times \vec c\,] = \left( \xi_\text{EDM} - \phi^\text{WF}_\text{rot} \right) \vec e_x + \frac{\chi_1}{2\sin \pi \nu_s}  \vec e_z \,.
\end{equation}
Thus the resonance strength squared can be written as a sum of two independent quadratic functions,
\begin{equation}
  \left(\epsilon^\text{EDM}\right)^2 = \frac{\psi_\text{WF}^{2}}{16\pi^2} \left[ \left(\xi_\text{EDM} -  \phi^\text{WF}_\text{rot}\right)^2 + \left(\frac{ \chi_1}{2\sin \pi \nu_s}\right)^2 \right]\, . \label{eq:EpsilonMap}
\end{equation}
where $\psi_\text{WF}$ is defined in Eq.\,(\ref{eq:psi-angle-in-WF}). 
			

\end{appendix}

\bibliographystyle{apsrev4-2}
\bibliography{/home/frank/Dokumente/Documents/LITERATUR/spintunemapping_23.01.2017}
\end{document}